\documentclass[a4paper,11pt]{article}
\usepackage{jheppub} % for details on the use of the package, please
\usepackage[T1]{fontenc} % if needed

\usepackage{enumitem}
\usepackage{multirow}

\title{\boldmath Testing predictions of discrete flavour and modular symmetries at DUNE and Hyper-K with JUNO constraints}

\author[a] {Debajyoti Dutta}
\author[b] {Srubabati Goswami}
\author[b,c,1] {Monal Kashav,\note{Corresponding author.}}
\author[b] {Ketan M. Patel}

\affiliation[a]{Department of Physics, Bhattadev University, Assam 781325, India}

\affiliation[b]{Theoretical Physics Division, Physical Research Laboratory, Navarangpura, Ahmedabad-380009, India}
\affiliation[c]{Department of Electronics and Communication Engineering, SR University, Warangal, Telangana, 506371}
\emailAdd{phy.debajyoti@bhattadevuniversity.ac.in}
\emailAdd{sruba@prl.res.in}
\emailAdd{monalkashav@gmail.com}
\emailAdd{ketan.hep@gmail.com}

\abstract{ We study fixed-column predictions of the lepton mixing matrix that arise from residual symmetries originating in a class of discrete flavour and modular symmetries. While the recent high-precision determination of $\sin^{2}\theta_{12}$ by JUNO already constrains part of these predictions, the remaining viable scenarios are primarily characterized by non-trivial correlations between $\sin^{2}\theta_{23}$ and the Dirac CP phase $\delta_{\rm CP}$, which are currently only weakly constrained. This motivates a detailed investigation of the sensitivities of next-generation long-baseline neutrino experiments to these scenarios. For the phenomenologically viable fixed-column predictions that can constitute a column of the lepton mixing matrix, we derive precise $\sin^{2}\theta_{23}$–$\delta_{\rm CP}$ correlations and use them to generate test-event samples, marginalising over the remaining oscillation parameters. We perform detailed simulations for DUNE and Hyper-K, presenting the allowed regions in the $\sin^{2}\theta_{23}$–$\delta_{\rm CP}$ plane. We also evaluate, as a function of experimental exposure (determined by the run time), the fraction of $\delta_{\rm CP}$ values for which these theoretical predictions can be ruled out at more than $3\sigma$ CL.
Our results show that the combined sensitivity of DUNE and Hyper-K provides a robust test of fixed-column lepton-mixing predictions.
}
\begin{document} 
\maketitle
\flushbottom

\section{Introduction}
Over the last two decades, neutrino oscillation experiments have established a detailed picture of lepton mixing, starting with atmospheric oscillations at Super-Kamiokande, followed by the solar sector measurements at SNO \cite{SNO2001} and KamLAND \cite{ KamLAND2003}, and the determination of a sizeable reactor angle $\theta_{13}$ by Double Chooz \cite{ DoubleChooz2012}, Daya Bay \cite{DayaBay2012}, and RENO \cite{RENO2012}. Consequently, the mass-squared differences $\Delta m^2_{21}$, $\Delta m^2_{31}$ and the mixing angles $\theta_{12}$, $\theta_{23}$, and $\theta_{13}$ in the PMNS matrix are now known with good precision. The main remaining unknowns are the neutrino mass ordering, the octant of $\theta_{23}$, and the Dirac CP phase $\delta_{\rm CP}$, with $\Delta m^2_{31}>0$ ($<0$) corresponding to normal (inverted) ordering. The measurement of a non-zero $\theta_{13}$ opened the door to probing the mass ordering and leptonic CP violation, ushering in the precision era. In this landscape, JUNO is particularly important: in addition to its sensitivity to the mass ordering \cite{Petcov:2001sy,Choubey:2003qx}, its baseline enables a high-precision determination of the solar mixing angle $\theta_{12}$ \cite{Bandyopadhyay:2003du,Bandyopadhyay:2004cp}. The recent 60-day JUNO data \cite{JUNO:2025gmd} have improved the global-fit precision of $\sin^{2}\theta_{12}$ by about a factor of $1.6$, yielding $0.3087 ^{+0.0069}_{-0.0066}$ and $\Delta m^{2}_{21}=(7.533^{+0.096}_{-0.10})\times10^{-5}\,\mathrm{eV}^{2}$ \cite{Capozzi:2025ovi,Esteban:2026phq}.

At the same time, the long baseline neutrino experiments T2K \cite{T2K:2021xwb,T2K:2023smv} and NO$\nu$A \cite{NOvA:2021nfi, NOvA:2025tmb} are taking data. The inferences from their data do not agree well if we assume normal ordering (NO) in neutrino masses. Their $1\sigma$ allowed regions do not overlap, reflecting the tension between the two experiments.  However, in the case of inverted ordering (IO), the two experiments give almost the same best-fit values, and their $1\sigma$ regions overlap. This means the datasets are statistically consistent under IO, and the combined fit therefore prefers IO as the overall best option \cite{Mikola:2024jnj,T2K:2025wet}. The main reason that the tension exists, for NO, can be ascribed to the existence of a hierarchy-$\delta_{CP}$ degeneracy \cite{Minakata:2003wq, Mena:2004sa}. In fact, after the precise determination of the mixing angle, $\theta_{13}$ though,  some components of the original eight-fold degeneracy \cite{Barger:2001yr} were resolved; it was pointed out in \cite{Ghosh:2015ena} that the remaining degeneracies can be understood in terms of a generalized hierarchy-octant-$\delta_{CP}$ degeneracy, which includes a hierarchy-$\delta_{CP}$, hierarchy-octant, and octant-$\delta_{CP}$ degeneracy \cite{ Gandhi:2004bj,Agarwalla:2013ju}. The use of a combined neutrino and antineutrino run can help to remove the octant degeneracies, thereby improving the precision of  determination of $\delta_{CP}$ as well \cite{Agarwalla:2013ju, Ghosh:2014dba,Ghosh:2015ena}. 
Recently the role of JUNO in removing hierarchy-$\delta_{CP}$ degeneracy has been studied in 
\cite{Goswami:2025wla}. The capability of  future experiments like DUNE and Hyper-K in  resolving  the above ambiguities has been investigated, for instance, in 
\cite{Nath:2015kjg,Ghosh:2014rna,Chakraborty:2017ccm,Goswami:2016auo}.

The leptonic mixing angles are significantly larger than those in the quark sector, motivating theoretical frameworks that aim to explain this disparity. Discrete flavour symmetries applied to the leptonic sector are particularly attractive, as they yield predictive mixing patterns testable by precision neutrino data \cite{Altarelli:2010gt,Ishimori:2010au,King:2013eh,King:2014nza,Petcov:2017ggy,Xing:2020ijf,Feruglio:2019ybq,Almumin:2022rml}. In this precision era, it is essential to examine how improved measurements constrain such predictions. Residual symmetries—specifically a $Z_2$ symmetry in the neutrino sector combined with a $Z_m$ symmetry ($m\geq3$) in the charged-lepton sector—strongly restrict the PMNS matrix $\mathrm{U}_{\rm PMNS}$ and lead to fixed-column predictions. The exact nature of these residual symmetries can be effectively constrained if they are assumed to originate from the discrete subgroups (DSG) of $U(3)$ (that also include DSG of $SU(3)$) or arise from a class of modular symmetries at a fixed point in moduli. The symmetry dictated fixed-column configurations lead to correlated predictions among the leptonic mixing angles $\theta_{12}$, $\theta_{23}$, $\theta_{13}$ and the Dirac CP-violating phase $\delta_{\rm CP}$. These correlations constitute distinctive phenomenological signatures that can be probed in neutrino oscillation experiments. The experimental prospects for probing such correlations have been investigated in a number of works in the context of present and future neutrino facilities \cite{deAdelhartToorop:2011nfg,Hanlon:2013ska,Hanlon:2014bga,Pasquini:2018udd,Srivastava:2017sno,Agarwalla:2017wct,Chatterjee:2017ilf,Petcov:2017ggy,Petcov:2018snn,Ballett:2016yod,Ballett:2015wia,Ballett:2014dua,Penedo:2017knr,Blennow:2020snb,Blennow:2020ncm}.

The recent JUNO measurement of $\theta_{12}$ represents a major advance in this programme. In fixed-column scenarios, the precise determination of $\sin^2\theta_{12}$ by JUNO directly constrains the first element of the predicted column given a well-measured range of $\sin^2\theta_{13}$. Following the JUNO results, several studies have examined the implications of the improved $\theta_{12}$ precision
\cite{Ding:2025dqd,Petcov:2025aci,Zhang:2025jnn,Ge:2025csr,Jiang:2025hvq,He:2025idv,Ding:2025dzc,Borah:2025vtn,Shang:2026qkh}. Most of these works primarily focus on assessing whether the predicted mixing patterns or fixed column predictions remain consistent with the JUNO-allowed range of $\theta_{12}$. However, the remaining entries of the fixed column depend on $\theta_{23}$ and $\delta_{\rm CP}$, which are still affected by sizeable experimental uncertainties. A detailed investigation of the correlated predictions from the residual symmetries in the $\theta_{23}$–$\delta_{\rm CP}$ plane, which constitute a defining feature of residual symmetry frameworks, has received comparatively less attention. Since such correlations provide a direct experimental handle to distinguish symmetry-driven scenarios from more generic phenomenological approaches, they deserve a dedicated and systematic study.

In this work, we focus on testing the correlations among the leptonic mixing parameters predicted by different discrete symmetry groups at the upcoming long-baseline neutrino experiments DUNE \cite{DUNE:2016hlj,DUNE:2016rla,DUNE:2021cuw} and Hyper-K \cite{Hyper-KamiokandeProto:2015xww,Hyper-Kamiokande:2016srs}. Specifically, we ask whether residual-symmetry–predicted correlations between $\theta_{23}$ and $\delta_{\rm CP}$ can be decisively tested and constrained by future long-baseline experiments, and how effectively DUNE and Hyper-K, individually and jointly, together with the anticipated precise measurement of $\sin^2\theta_{12}$ from JUNO can achieve this. Owing to their high-intensity neutrino beams, large detector volumes, and long baselines, both experiments are expected to achieve unprecedented precision in the determination of neutrino oscillation parameters. In particular, the combination of enhanced statistics and complementary experimental configurations enables DUNE and Hyper-K to efficiently probe, scrutinize, and potentially exclude symmetry-driven correlations in the leptonic sector. Our analysis is performed for fixed-column predictions arising from DSGs of $U(3)$,  $SU(3)$ and modular symmetries that are compatible with current global
neutrino oscillation data.

The structure of the paper is as follows. In Sec.~\ref{sec:residual}, we briefly review the mechanism by which fixed-column predictions emerge from residual symmetries. In Sec.~\ref{sec:correlations}, we discuss the resulting correlations among the leptonic mixing parameters. The experimental setup, simulation details, and analysis procedure for DUNE and Hyper-K are described in Sec.~\ref{sec:experiment}. Our simulation results and their implications are presented in Sec.~\ref{sec:results}. Finally, we summarize our findings and present our conclusions in Sec.~\ref{sec:summary}.

\section{Residual symmetries and fixed column predictions}
\label{sec:residual}
The basic mechanism that leads to group-theoretical predictions for a column of the leptonic mixing matrix, $U_{\rm PMNS}$, is as follows. It is postulated that a global discrete symmetry acts non-trivially on left-chiral leptons which subsequently leads to residual symmetries $G_\nu$ and $G_l$ that constrain the neutrino and charged-lepton mass matrices, respectively. Because the symmetry generators commute with the hermitian square of the corresponding mass matrices, the leptonic mixing matrix $U_{\rm PMNS}$ is directly related to the diagonalisation of these generators \cite{Lam:2007qc,Lam:2008rs,Lam:2008sh,Lam:2012ga,Lam:2011ag}.  While predicting entire $|U_{\rm PMNS}|$ consistent within the current experimental bound requires very large flavour groups and is therefore poorly motivated, a more viable strategy is to predict only a single column, namely $c_0$, of $U_{\rm PMNS}$. This is achieved by taking $G_\nu=Z_2$ and $G_l = Z_m$, with $m\geq3$, for which a unique eigenvector of the neutrino symmetry fixes one column of the mixing matrix up to permutations, reflecting the absence of charged-lepton mass predictions. The technical details and explicit derivations underlying the above discussion are presented in Appendix \ref{app:DFS_background}.

Following the above strategy and a systematic analytical and numerical scan of a large category of groups, several candidates for fixed column $|c_0|$ have been obtained. The investigated groups include (a) DSG of $SU(3)$ with order $< 512$ \cite{Grimus:2010ak,Grimus:2011fk,Grimus:2013apa}, (b) DSG of $U(3)$ which are not subgroups of $SU(3)$ with order $<512$ \cite{Ludl:2010bj,Joshipura:2014pqa,Jurciukonis:2017mjp}, and (c) finite modular groups $\Gamma_N$ with $N\leq 16$ \cite{Kashav:2024lkr}. Category (a) comprises the group series $\Delta(3n^{2})$, $\Delta(6n^{2})$, and other groups belonging to the $\Sigma(m)$ class. DSGs of $U(3)$ that are not subgroups of $SU(3)$, when considered as candidates for flavour symmetries, are more restrictive and can impose additional constraints on neutrino masses \cite{Joshipura:2013pga}. This category therefore also includes the group series $\Delta(3n^{2},m)$ and $\Delta(6n^{2},m)$, along with other groups listed in \cite{Ludl:2010bj}. In addition, the modular symmetry group $\Gamma_{N}$ may act as a residual symmetry at fixed points of the moduli space. Overall, we restrict our attention to groups that possess at least one faithful three-dimensional irreducible representation, allowing the three generations of lepton doublets to be consistently embedded.

%%%%%%%%%%%%%%%%%%%%%%%%%%%%%%%%%%%%%%%%%%
\begin{table}[!htbp]
\centering
\footnotesize
\setlength{\tabcolsep}{4pt}
\renewcommand{\arraystretch}{1.2}

\begin{tabular}{%
l@{\hspace{18pt}} c@{\hspace{26pt}}
p{3cm}@{\hspace{26pt}}
p{3cm}
c
}
\hline\hline
$|c_0|^2$
& Identification 
& DSG $\mathrm{SU}(3)$
& DSG $\mathrm{U}(3)$
& $\Gamma_N$ \\
\hline\hline

$(2/3,\;1/6,\;1/6)$
& $C1[1]$
& $\Delta(6\times2^2)$, $\Delta(6\times4^2)$, $\Delta(6\times6^2)$, $\Delta(6\times8^2)$,
$PSL(2,7)$, $\Sigma(168)$, $\Sigma(360\phi)$
& ---
& $\Gamma_{4,7,16}$ \\
\hline

% $(0.724,\;0.138,\;0.138)$
% & $C1[2]$
% & ---
% & ---
% & $\Gamma_5$ \\
% \hline

% ---------- C1[2] ----------
$(0.6594,\;0.2303,\;0.1103)$
& $C1[2]$
& \multirow{2}{*}{\centering $\Delta(6\times5^2)$}
& \multirow{2}{*}{\centering ---}
& \multirow{2}{*}{\centering ---} \\
$(0.6594,\;0.1103,\;0.2303)$
& $C1'[2]$
& 
& 
&  \\
\hline

% ---------- C1[3] ----------
$(0.6629,\;0.2116,\;0.1255)$
& $C1[3]$
& \multirow{2}{*}{\centering $\Delta(6\times7^2)$}
& \multirow{2}{*}{\centering ---}
& \multirow{2}{*}{\centering ---} \\
$(0.6629,\;0.1255,\;0.2116)$
& $C1'[3]$
& 
& 
&  \\
\hline

% ---------- C1[4] ----------
$(0.6640,\;0.2045,\;0.1315)$
& $C1[4]$
& \multirow{2}{*}{\centering $PSL(2,7)$, $\Sigma(168)$}
& \multirow{2}{*}{\centering ---}
& \multirow{2}{*}{\centering $\Gamma_7$} \\
$(0.6640,\;0.1315,\;0.2045)$
& $C1'[4]$
& 
& 
&  \\
\hline

% ---------- C1[5] ----------
$(0.6553,\;0.2471,\;0.0976)$
& $C1[5]$
& \multirow{2}{*}{\centering $\Delta(6\times8^2)$}
& \multirow{2}{*}{\centering ---}
& \multirow{2}{*}{\centering $\Gamma_{16}$} \\
$(0.6553,\;0.0976,\;0.2471)$
& $C1'[5]$
& 
& 
&  \\
\hline

% ---------- C1[6] ----------
$(0.6466,\;0.2755,\;0.078)$
& $C1[6]$
& \multirow{3}{*}{\centering
$\Delta(6\times9^2)$, $D^1_{9,3}$, $\Sigma(216\phi)$}
& 
$[[162,10]]$,$[[162,12]]$,
$[[486,26]]$,$[[486,28]]$,
$[[486,125]]$
& \multirow{2}{*}{\centering ---} \\
$(0.6466,\;0.078,\;0.2755)$
& $C1'[6]$
& 
& 
&  \\
\hline

% ---------- C1[7] ----------
$(0.7057,\;0.2500,\;0.0443)$
& $C1[7]$
& \multirow{2}{*}{\centering $PSL(2,7)$, $\Sigma(168)$}
& \multirow{2}{*}{\centering ---}
& \multirow{2}{*}{\centering ---} \\
$(0.7057,\;0.0443,\;0.2500)$
& $C1'[7]$
& 
& 
&  \\
\hline
$(1/3,\;1/3,\;1/3)$
& $C2[1]$
& $\Delta(3n^2)$, $\Delta(6\times2^2)$, $\Delta(6\times4^2)$, $\Delta(6\times6^2)$, $\Delta(6\times8^2)$,
$A_5$, $\Sigma(60)$, $PSL(2,7)$, $\Sigma(168)$, $\Sigma(360\phi)$
& $S_4(2)$, $S_4(3)$, $S_4(5)$, $\Delta(6\times4^2,3)$
& $\Gamma_{3,5,7}$ \\
\hline
$(0.2764,\;0.3618,\;0.3618)$
& $C2[2]$
& $A_5$, $\Sigma(60)$, $\Sigma(360\phi)$
& ---
& $\Gamma_5$ \\
\hline\hline
\end{tabular}
\caption{Viable fixed-column predictions ($|c_0|^2$) for leptonic mixing realized by discrete symmetry groups, with listed discrete subgroups of  $\mathrm{SU}(3)$, $\mathrm{U}(3)$, and modular groups $\Gamma_N$ corresponding to each $|c_{0}|^2$. Labels $C1[n]$ ($C2[n]$) denote candidates associated with the first (second) column of $U_{\rm PMNS}$ within the $3\sigma$ NuFIT~6.0 ranges~\cite{Esteban:2024eli}. For each $C1[n]$, a permuted counterpart $C1'[n]$ is obtained by interchanging the second and third entries of $|c_0|^2$.}
\label{tab:1}
\end{table}
%%%%%%%%%%%%%%%%%%%%%%%%%%%%%%%%%%%%%%%%%%
We collect in Table \ref{tab:1} all these candidates which can be viably identified with either of the three columns of $U_{\rm PMNS}$ within their $3 \sigma$ range provided by the pre-JUNO global analysis results NuFIT 6.0\footnote{For consistency, we use the global analysis NuFIT 6.0 and 6.1 results that do not incorporate atmospheric neutrino data provided by the Super-Kamiokande and IceCube collaborations.} \cite{Esteban:2024eli}. The latter is reproduced below for convenience.
\begin{equation}
|U_{\rm PMNS}|^2_{3 \sigma} = 
\begin{pmatrix}
0.6416 \,\rightarrow\, 0.7090 & 0.2694 \,\rightarrow\, 0.3364 & 0.0202 \,\rightarrow\, 0.0240 \\
0.0615 \,\rightarrow\, 0.2550 & 0.2237 \,\rightarrow\, 0.4651 & 0.4212 \,\rightarrow\, 0.5837 \\
0.0729 \,\rightarrow\, 0.2714 & 0.2333 \,\rightarrow\, 0.4761 & 0.3944 \,\rightarrow\, 0.5565
\end{pmatrix}.
\end{equation}
We subsequently assess the impact of the updated global analysis results from NuFIT 6.1 \cite{Esteban:2026phq} on these predictions. Some noteworthy aspects about the candidate solutions listed in Table \ref{tab:1} are the following. 
\begin{itemize}
    \item None of the groups considered above can reproduce the third column of $|U_{\rm PMNS}|$. This reiterates the fact that reproducing $|U_{e3}|$ consistent with the present observations using small-sized discrete groups is impossible, as mentioned earlier.
    \item Only two candidate predictions exist for the second column of $|U_{\rm PMNS}|$. One of them is the so-called tri-maximal mixing scenario, which can emerge from small groups like $A_4$. In contrast to this, several viable candidate predictions can be identified as the first column.
    \item The solutions  $C1[1]$, $C2[1]$ and $C2[2]$ have identical magnitudes of the second and third elements. These cases, therefore, come under the partial $\mu-\tau$ reflection symmetry considered earlier in a similar kind of analysis \cite{Chakraborty:2018dew}.
    \item  The columns $C1[n]$ with $n\geq 3$ have distinct second and third elements. The permutations between the two also lead to a viable column, which we identify as $C1^\prime[n]$. These are the cases corresponding to broken $\mu-\tau$ reflection symmetry. 
\end{itemize}

\section{Correlations among the leptonic mixing parameters}
\label{sec:correlations}
The predictions listed in Table \ref{tab:1} lead to specific correlations among certain leptonic mixing parameters. It can be seen that the first entry of $C2[n]$ or $C1[n]$ can only be identified with the first element of the corresponding column in the above $|U_{\rm PMNS}|^2$. Hence, this fixes the correlation between $\theta_{12}$ and $\theta_{13}$ when compared with the standard parametrisation. Explicitly, 
\begin{equation} \label{Eqn:7}
\cos^2 \theta_{13} \cos^2 \theta_{12} = C1[n]_1\, \quad {\rm or} \quad \cos^2 \theta_{13} \sin^2 \theta_{12} = C2[n]_1\,
\end{equation}
for each case. Using the $3 \sigma$ range of $\theta_{13}$, the above correlations can directly be converted to find the corresponding range of $\sin^2 \theta_{12}$. We carry out this exercise for the listed solutions in the context of both NuFIT 6.0 and 6.1 and present the results in Table~\ref{tab:2}. We also show the compatibility of the model predictions for $\theta_{12}$ with the corresponding global-fit results. The $3\sigma$ ranges of the columns of the leptonic mixing matrix from the updated global analysis NuFIT 6.1, which includes the JUNO data, are given as,
\begin{equation} \label{UPMNS_nufit61}
|U_{\rm PMNS}|^2 =
\begin{pmatrix}
0.6561 \to 0.6956 & 0.2830 \to 0.3226 & 0.0207 \to 0.0243 \\
0.0666 \to 0.2520 & 0.2294 \to 0.4624 & 0.4225 \to 0.5746 \\
0.0751 \to 0.2621 & 0.2362 \to 0.4679 & 0.4032 \to 0.5550
\end{pmatrix}\,.
\end{equation}

%%%%%%%%%%%%%%%%%%%%%%%%%%%%%%%%%%%%%%%%%%%%%%%%%5555
\begin{table}[!htbp]
\centering
\renewcommand{\arraystretch}{1.2}
\begin{tabular}{c@{\hspace{25pt}} c@{\hspace{25pt}} c@{\hspace{25pt}} c@{\hspace{25pt}} c@{\hspace{25pt}} c}
\hline\hline
Identification
& $\sin^2\theta_{12}$ range 
& NuFIT~6.0
& $\sin^2\theta_{12}$ range 
& NuFIT~6.1 \\
\hline\hline
$C1[1]$ & $(0.3170,\,0.3196)$ & $\checkmark$ & $(0.3236,\,0.3260)$ & $\checkmark$  \\

$C1[2]$, $C1'[2]$ & $(0.3249,\,0.3275)$ & $\checkmark$ & $(0.3244,\,0.3267)$  & $\checkmark$ \\

$C1[3]$, $C1'[3]$ & $(0.3211,\,0.3233)$ & $\checkmark$ & $(0.3207,\,0.3234)$ & $\checkmark$ \\

$C1[4]$, $C1'[4]$ & $(0.3200,\,0.3220)$ & $\checkmark$ & $(0.3197,\,0.3220)$  & $\checkmark$ \\

$C1[5]$, $C1'[5]$ & $(0.3289,\,0.3312)$ & $\checkmark$ & $(0.3277,\,0.3302)$ & $\checkmark$ \\

$C1[6]$, $C1'[6]$ & $(0.3376,\,0.3400)$ & $\checkmark$ & $(0.3407,\,0.3433)$ & $\times$  \\

$C1[7]$, $C1'[7]$ & $(0.2772,\,0.2797)$ & $\checkmark$ & $(0.2803,\,0.2830)$ & $\times$  \\

$C2[1]$ & $(0.3402,\,0.3415)$ & $\checkmark$ & $(0.3369,\,0.3381)$ & $\times$  \\

$C2[2]$ & $(0.2881,\,0.2902)$ & $\checkmark$ & $(0.2822,\,0.2833)$ & $\times$  \\

\hline\hline
\end{tabular}
\caption{Compatibility of squared PMNS matrix elements with NuFIT~6.0 and NuFIT~6.1 results. The values of $\sin^2\theta_{12}$ given in the second and fourth columns are computed using Eq. (\ref{Eqn:7}) and the $3\sigma$ range of $\sin^2\theta_{13}$ from NuFIT 6.0 and 6.1, respectively.} Checkmarks ($\checkmark$) and crosses ($\times$) indicate allowed and disfavoured ranges of $\sin^2\theta_{12}$, respectively.
\label{tab:2}
\end{table}
%%%%%%%%%%%%%%%%%%%%%%%%%%%%%%%

Recent analyses of lepton flavour models based on the non-Abelian discrete groups $A_{5}$, $\Sigma(168)$ and $\Delta(6n^{2})$, supplemented by generalised CP symmetry, show that residual symmetries lead to predictive one- and two-parameter mixing patterns \cite{Ding:2025dqd}. It was reported that the recent JUNO determination of $\sin^{2}\theta_{12}$ significantly reduces the viable parameter space for one-parameter scenarios: several well-known patterns, such as $C2[1]$ i.e. tri-maximal(TM)—characterized by equal mixing among the three lepton flavours in one column of the PMNS matrix \cite{Grimus:2008tt,Albright:2010ap} and the golden-ratio (GR) mixing pattern—where $\theta_{23}$ is maximal, $\theta_{13}=0$, and the solar angle $\theta_{12}$ is determined by the golden ratio $\phi_g=(1+\sqrt{5})/2$\cite{Datta:2003qg,Kajiyama:2007gx}—are strongly disfavoured by current data. In contrast, the $C1[1]$ pattern corresponding to tri-large (TL) mixing, characterized by the mixing fractions $(2/3,,1/6,,1/6)$ for a PMNS column \cite{Joshipura:2016quv} remains compatible with present neutrino oscillation constraints. Other recent related studies on the status and compatibility of neutrino mixing patterns in light of the JUNO results can be found in Refs.~\cite{Petcov:2025aci,Zhang:2025jnn,Ge:2025csr,Jiang:2025hvq,He:2025idv,Ding:2025dzc}. The viable fixed-column predictions arising from the aforementioned discrete and modular symmetries that remain consistent with the JUNO measurement, as well as the latest global fit, are listed in Table~\ref{tab:2}. These findings underscore the strong discriminatory power of precision measurements of $\theta_{12}$ over one-parameter(fixed column) residual-symmetry constructions and motivate a detailed study of the surviving correlations involving $\theta_{23}$ and the Dirac CP phase $\delta_{\rm CP}$ due to uncertainty in their measurement.

The other elements of the fixed-column solutions can be used to obtain correlations between the Dirac CP phase and mixing angles. For example, one finds by comparing $|U_{\mu 1}|$ with either of $|C1[n]_2|$ or $|C1[n]_3|$, 
\begin{equation} \label{Eqn:8}
\cos\delta =
\frac{|C1[n]|_{2,3} - s_{12}^{2} c_{23}^{2}
  - c_{12}^{2} s_{13}^{2} s_{23}^{2}}{2\, s_{13} c_{12} s_{12} c_{23} s_{23}}\,
\end{equation}
where $c_{ij} = \cos \theta_{ij}$ and $s_{ij}=\sin \theta_{ij}$. Similarly, by comparing $|U_{\tau 1}|$ with $|C1[n]_3|$ or $|C1[n]_2|$, one finds  
\begin{equation} \label{Eqn:9}
\cos\delta =
\frac{-|C1[n]|_{3,2} + s_{12}^{2} s_{23}^{2}
  + c_{12}^{2} s_{13}^{2} c_{23}^{2}}{2\, s_{13} c_{12} s_{12} c_{23} s_{23}}
\end{equation}
As a special case when $|C1[n]_2| = |C1[n]_3|$, i.e. realising the $\mu$-$\tau$ reflection symmetry for this column, the last two equations reduce to 
\begin{equation} \label{Eqn:10}
\cos\delta =
\frac{(s_{23}^2 -c_{23}^2)(s_{12}^2-c_{12}^2 s_{13}^2)}{4\, s_{13} c_{12} s_{12} c_{23} s_{23}}\,.
\end{equation}

Similar correlations can be obtained for the solutions that are identified with the second column. It can be seen from Table \ref{tab:1} that all the viable solutions in this case already satisfy the partial $\mu$-$\tau$ reflection symmetry. As a result, a simple correlation
\begin{equation}  \label{Eqn:11}
\cos\delta =
\frac{(c_{23}^2 -s_{23}^2)(c_{12}^2-s_{12}^2 s_{13}^2)}{4\, s_{12} c_{12} s_{13} c_{23} s_{23}}\,,
\end{equation}
exists for all the solutions. The correlations among the leptonic mixing parameters derived above from the fixed-column predictions provide constrained ranges for 
$\sin^2\theta_{12}$ and the Dirac CP phase $\delta_{\rm CP}$, which will be used to generate test events in our experiment simulations in the next section. All other oscillation parameters will be marginalised over their allowed ranges to focus on the sensitivity of the simulated data to the predicted correlations in the $\theta_{23}$--$\delta_{\rm CP}$ plane.

\section{Experimental and Simulation details  }\label{sec:experiment}

\subsection{Experimental Setups: DUNE and Hyper-K}

\paragraph{DUNE:}
The Deep Underground Neutrino Experiment (DUNE)~\cite{DUNE:2016hlj, DUNE:2016rla, DUNE:2021cuw, DUNE:2015lol, DUNE:2020fgq, LBNE:2013dhi} is a next-generation long-baseline neutrino oscillation experiment under construction in the United States. It will utilise a high-intensity neutrino beam produced at the Long-Baseline Neutrino Facility (LBNF) at Fermilab and directed toward massive detectors located deep underground in South Dakota. Our analysis is based on a DUNE configuration corresponding to an integrated exposure of $624~\text{kt}\cdot\text{MW}\cdot\text{yr}$, achieved through $6.5$ years of data taking in each of the neutrino (FHC) and antineutrino (RHC) modes with a $40$-kt fiducial mass detector and a $120$-GeV, $1.2$-MW proton beam. This configuration effectively reproduces 10 years of operation under the nominal staging scenario discussed in \cite{DUNE:2020jqi}. After traversing a baseline of $1300~\mathrm{km}$, the neutrino beam is observed at the Far Detector (FD) located at the Sanford Underground Research Facility (SURF) in South Dakota. The FD consists of four liquid argon time projection chamber (LArTPC) modules with a total fiducial mass of about $40~\mathrm{kt}$~\cite{DUNE:2021cuw, DUNE:2020jqi}. The LArTPC technology provides precise three-dimensional event reconstruction, excellent calorimetric energy resolution, and efficient particle identification. In the energy range $0.5$--$4$~GeV relevant for oscillation measurements, the neutrino energy resolution at the FD is approximately $15$--$20\%$, depending on the lepton flavour and reconstruction method. A constant matter density of $2.848~\text{g/cm}^3$ with a $2\%$ uncertainty is assumed throughout this analysis. For DUNE, the energy resolution is implemented using a detector-specific smearing matrix that accounts for reconstruction effects.

\begin{figure}
    \centering
    \includegraphics[width=0.98\linewidth]{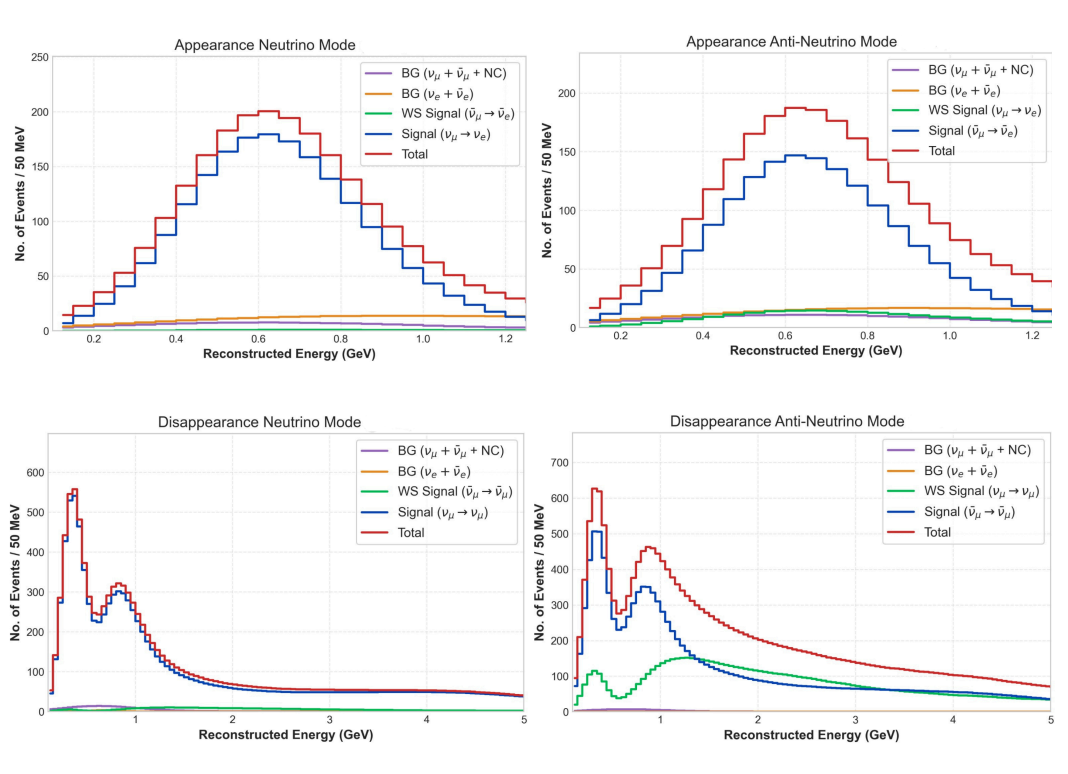}
    \caption{Reconstructed energy distributions for appearance (top panels) and disappearance (bottom panels) channels in neutrino (left) and anti-neutrino (right) modes. The contributions from signal, wrong-sign (WS) signal, and different background components are shown separately, along with their total. The appearance channels are dominated by the $\nu_\mu \rightarrow \nu_e$ ($\bar{\nu}_\mu \rightarrow \bar{\nu}_e$) transitions, while the disappearance channels correspond to $\nu_{\mu} \rightarrow \nu_{\mu}$ ($\bar{\nu}_\mu \rightarrow \bar{\nu}_\mu$) oscillations. The distributions are presented as a function of reconstructed energy with a bin width of 50 MeV.}
    \label{fig:events}
\end{figure}
\bigskip
\paragraph{Hyper-K:}
The Hyper-Kamiokande (Hyper-K) experiment~\cite{Hyper-KamiokandeProto:2015xww, Hyper-Kamiokande:2016srs, Hyper-Kamiokande:2018ofw, Hyper-Kamiokande:2025fci} is a proposed long-baseline neutrino oscillation experiment with a baseline of $295~\mathrm{km}$. An intense neutrino beam will be produced at the Japan Proton Accelerator Research Complex (J-PARC) and directed towards the large water-Cherenkov Hyper-Kamiokande detector. The J-PARC beam is expected to begin operation at a power of $750~\mathrm{kW}$, with a planned upgrade to $1.3~\mathrm{MW}$, delivering approximately $2.7 \times 10^{22}$ protons on target per year.

Hyper-Kamiokande is a next-generation water Cherenkov detector consisting of a single cylindrical module with a fiducial mass of $187~\mathrm{kt}$. The detector will be located $2.5^\circ$ off-axis with respect to the J-PARC beam, optimising its sensitivity to neutrino oscillation parameters. It will enable high-precision measurements of neutrino oscillations using accelerator, atmospheric, and solar neutrinos, and will probe CP violation in the lepton sector. In addition, Hyper-K will search for nucleon decay and detect astrophysical neutrinos, thereby advancing our understanding of fundamental physics and the Universe.

For the simulations presented in this work, we assume a total runtime of $10$ years, divided into $2.5$ years in neutrino mode and $7.5$ years in antineutrino mode, corresponding to a $1{:}3$ running ratio chosen to achieve comparable statistical sensitivity in both channels~\cite{Hyper-Kamiokande:2018ofw}. The appearance and disappearance event spectra in both $\nu$ and $\bar{\nu}$ modes are matched to those reported in Ref.~\cite{Hyper-Kamiokande:2018ofw}. The background contributions are also reproduced accordingly, and the resulting plots are shown in Fig.~\ref{fig:events}.

The energy resolution is modeled using a channel-dependent parameterization of the form
\[
\sigma(E) = \alpha E + \beta \sqrt{E} + \gamma \, .
\]
The parameters $\alpha$, $\beta$, and $\gamma$ are chosen separately for different interaction channels (electron/muon and QE/non-QE) in order to account for their distinct reconstruction characteristics. Their values are selected within realistic ranges and tuned to reproduce the appearance and disappearance event spectra, as well as background distributions, reported by the Hyper-Kamiokande collaboration in ~\cite{Hyper-Kamiokande:2018ofw}.

\subsection{Details of the $\Delta \chi^{2}$  Analysis} 
\label{chi2}

To test the model predictions, we perform a $\chi^2$ analysis using GLoBES \cite{Huber:2004ka, HUBER2007439} where $\Delta \chi^{2}$ is defined as
\begin{equation} \label{Eqn:12}
\begin{split}
    \Delta\chi^{2}(p^{\text{true}}) = \min_{p^{\text{test}},\eta} \Bigg[ & 2\sum_{i,j,k}^{} \left\{ N_{ijk}^{\text{test}}(p^{\text{test}};\eta) - N_{ijk}^{\text{true}}(p^{\text{true}}) + N_{ijk}^{\text{true}}(p^{\text{true}}) \ln\frac{N_{ijk}^{\text{true}}(p^{\text{true}})}{N_{ijk}^{\text{test}}(p^{\text{test}};\eta)} \right\} \\
    & + \sum_{l} \frac{(p_{l}^{\text{true}} - p_{l}^{\text{test}})^{2}}{\sigma_{p_{l}}^{2}} + \sum_{m} \frac{\eta_{m}^{2}}{\sigma_{\eta_m}^{2}} \Bigg],
\end{split}
\end{equation}
Here, $N^{\text{true}}$ denotes the simulated event rates evaluated at the true
values of the oscillation parameters and is treated as the experimental data, while $N^{\text{test}}$ represents the predicted event rates obtained in the fit
for a given test hypothesis. The quantities $p^{\text{true}}$ and
$p^{\text{test}}$ correspond to the true and test sets of oscillation parameters, respectively. The indices $i$, $j$, and $k$ run over the reconstructed energy bins, oscillation channels, and running modes (neutrino or antineutrino).\\

For the DUNE analysis, a uniform energy binning of $0.125~\mathrm{GeV}$ is adopted up to $8~\mathrm{GeV}$. For energies above $8~\mathrm{GeV}$ and extending up to $120~\mathrm{GeV}$, variable bin widths of $1~\mathrm{GeV}$, $2~\mathrm{GeV}$, and $10~\mathrm{GeV}$ are used. In the case of Hyper-K, the analysis employs uniform energy bins with a bin width of $0.05~\mathrm{GeV}$, covering the shown energy range in Fig. \ref{fig:events}.\\

We use the \textit{method of pulls} \cite{Huber:2002mx, Gandhi:2007td} to incorporate the systematics in our analysis. The last term in Eq.~\eqref{Eqn:12} is the pull term through which the nuisance parameters $\eta_{m}$ encode the
dependence of the predicted event rates on various sources of systematic
uncertainties. A complete list of the systematic uncertainties considered in this
analysis is summarised in Table~\ref{tab:uncertainity}. 

\begin{table}[!h]
\centering
\renewcommand{\arraystretch}{1.1}
\begin{tabular}{|c|c|c|}
\hline
\multirow{2}{*}{Channel} &
\multicolumn{2}{c|}{Normalization uncertainty} \\ \cline{2-3}
& Signal & Background \\
\hline \hline
\multicolumn{3}{|c|}{\textbf{DUNE}} \\
\hline
$\nu_e \;(\bar{\nu}_e)$ appearance
& $2\% \; (2\%)$ & $5\% \; (5\%)$ \\
$\nu_\mu \;(\bar{\nu}_\mu)$ disappearance
& $5\% \; (5\%)$ & $5\% \; (5\%)$ \\
\hline
\multicolumn{3}{|c|}{\textbf{Hyper-K}} \\
\hline
$\nu_e \;(\bar{\nu}_e)$ appearance
& $3.2\% \; (3.9\%)$ & $5\% \; (5\%)$ \\
$\nu_\mu \;(\bar{\nu}_\mu)$ disappearance
& $3.6\% \; (3.6\%)$ & $5\% \; (5\%)$ \\
\hline
\end{tabular}
\caption{Normalisation uncertainties ($\sigma_{\eta}$) associated with signal and background event rates for the oscillation channels considered in this analysis. The values in parentheses correspond to the antineutrino mode. For DUNE, additional background components such as $\nu_\tau$ and NC backgrounds are included with larger normalization uncertainties (20\% and 10\%, respectively).}
\label{tab:uncertainity}
\end{table}

In this work, we explicitly study the impact of priors on
$\sin^{2}\theta_{12}$. The anticipated high-precision measurement of $\theta_{12}$ from the
JUNO experiment is incorporated through a Gaussian prior on
$\sin^{2}\theta_{12}$, defined as
\begin{equation}  \label{Eqn:13}
\chi^{2}_{\text{JUNO}}(\sin^{2}\theta_{12}) =
\left(
\frac{\sin^{2}\theta_{12}^{\text{fit}} - \sin^{2}\theta_{12}^{\text{bf}}}
{0.0087}
\right)^{2} .
\end{equation}
Here, $\chi^{2}_{\text{JUNO}}(\sin^{2}\theta_{12})$ represents the Gaussian $\chi^2$ contribution from the expected precision measurement of the solar mixing angle $\theta_{12}$ at JUNO. Here, $\sin^{2}\theta_{12}^{\text{fit}}$ denotes the test value used in the fit, while $\sin^{2}\theta_{12}^{\text{bf}}$ corresponds to the assumed best-fit (true) value taken from the global-fit data. The quantity $0.0087$ represents the $1\sigma$ uncertainty on $\sin^{2}\theta_{12}$ measured by JUNO. Thus, this term quantifies the deviation of the fit value from the assumed true value in units of the expected experimental precision.\\
The corresponding prior term then enters the definition of $\Delta \chi^{2}$ in
Eq.~\eqref{Eqn:12}, which is subsequently minimised to obtain the
minimum value, $\chi^{2}_{\mathrm{min}}$.

\subsection{Details of the simulation}
\begin{table*}[!t]
    \centering 
    \renewcommand{\arraystretch}{1.3}
    \begin{tabular}{|c|c|c|}
    \hline 
    \rule{0pt}{12pt} Oscillation Parameter & Values for which data is generated &  Values in the fit\tabularnewline
    \hline \hline
    \rule{0pt}{12pt} $\sin^2\theta_{12}$       & 0.307               &  As per the model constraints  \tabularnewline \hline 
    \rule{0pt}{12pt} $\sin^2\theta_{13}$       & 0.02195                & within 3$\sigma$ allowed range \tabularnewline \hline 
    \rule{0pt}{12pt} $\theta_{23}$       & $41.0^{\circ}-50.5^{\circ}$               & within 3$\sigma$ allowed range\tabularnewline \hline 
    \rule{0pt}{12pt} $\delta_{\rm CP}$       & $0^{\circ}-360^{\circ}$                 & As per the model \tabularnewline \hline 
    \rule{0pt}{12pt} $\Delta m_{21}^{2}$ & $7.49\times10^{-5}\rm~eV^{2}$ & Fixed  \tabularnewline \hline
    \renewcommand{\arraystretch}{1.0}
    \rule{0pt}{12pt} $\Delta m_{31}^{2}$ &  $2.53\times10^{-3}\rm~eV^{2}$& $ [2.46,2.60] \times10^{-3}\rm~eV^{2}$\tabularnewline  
    \hline 
    \end{tabular}\\
    \vspace{0.2cm}
    \caption{
Summary of the neutrino oscillation parameters used in the analysis.
The table lists the benchmark (true) values adopted for event generation and the corresponding parameter ranges or conditions used in the fit and marginalisation procedure.
The values and allowed ranges are chosen in accordance with the NuFIT~6.1 global analysis~\cite{Esteban:2024eli}, while model-dependent constraints are imposed where applicable.
}
\label{tab:marginalization_val}
\end{table*} 
 The details of the simulations carried out in this work are as follows:
\begin{itemize}
\item The true values of the oscillation parameters $\theta_{12}$, $\theta_{13}$,
$\Delta m^{2}_{21}$, and $\Delta m^{2}_{31}$ are fixed in the simulated data at the
benchmark values listed in Table~\ref{tab:marginalization_val}.
The parameters $\theta_{23}$ and $\delta_{\rm CP}$ are allowed to vary over their
current $3\sigma$ ranges from NuFIT 6.1  ~\cite{Esteban:2026phq}.

\item In the fit, we marginalise over $|\Delta m^{2}_{31}|$ and $\theta_{23}$
within their respective $3\sigma$ allowed intervals.
For each value of $\sin^{2}\theta_{13}$ permitted by the NuFIT~6.1 $3\sigma$ range,
we determine the corresponding value of $\sin^{2}\theta_{12}$ as dictated by the
model-specific correlations.
Using these three mixing angles, the Dirac CP phase $\delta_{\rm CP}$ is then
constrained through the predicted value of $\cos\delta_{\rm CP}$ (see, Section \ref{sec:correlations}).

\item For each choice of the true oscillation parameters, the theoretical predictions
are compared with the simulated data by minimizing $\chi^{2}$ function with
respect to the relevant test parameters and nuisance parameters.
By scanning over the full allowed ranges of the true values of $\theta_{23}$ and
$\delta_{\rm CP}$, we obtain the corresponding $\chi^{2}_{\min}$ at each point in
parameter space.
This procedure allows us to construct the allowed regions in the
$\theta_{23}$--$\delta_{\rm CP}$ plane shown in the results section, based on fixed
$\Delta\chi^{2}$ criteria. The explicit form of the $\chi^{2}$ function adopted in this work is presented in the next subsection.

\item We have checked that  our analysis for both normal ordering (NO) and inverted ordering (IO) and found that the results do not change significantly. The main physics conclusions of the work remain essentially unchanged for both mass
orderings. Hence, we present the results assuming normal mass ordering as the true mass ordering. 

\end{itemize}

\section{Results} \label{sec:results}
The question that we want to answer through this work is: can the correlations between $\theta_{23}$ and $\delta_{\rm CP}$ predicted by the residual symmetries be decisively tested and constrained by future long-baseline experiments? If so, how effectively can DUNE and Hyper-K, both individually and in combination, together with the anticipated precision measurement of $\sin^2\theta_{12}$ from JUNO, accomplish this task?

To address these questions, we perform a detailed $\chi^2$ analysis as stated above. Including both appearance and disappearance channels in neutrino and anti-neutrino mode, we simulate prospective data for DUNE and Hyper-K and compute the corresponding $\chi^2$ for different assumed true values of $\sin^2\theta_{23}$ and $\delta_{\rm CP}$.

\subsection{Analysis of the allowed regions in $\theta_{23}-\delta_{\rm CP}$ plane}
Despite some of the fixed column predictions having tension with global-fit results due to their predictions for $\theta_{12}$ (see Table~\ref{tab:2}), we analyse their predicted correlations between $\sin^2\theta_{23}$--$\delta_{CP}$. Specific perturbations to this fixed-column prediction may modify their $\theta_{12}$ value, leaving one of the second or third entries of the column unchanged see for example \cite{deMedeirosVarzielas:2025byb}. In the present section, we, however, concentrate on simulation results for fixed-column predictions consistent with the NuFIT~6.1, while disfavored cases are discussed in Appendix~\ref{app:plots}. Figures~\ref{fig1}–\ref{fig5}, together with Figs.~\ref{fig8}, \ref{fig9}, and \ref{fig10}, illustrate the allowed regions in the true $\sin^2\theta_{23}$--$\delta_{\rm CP}$ parameter space corresponding to various residual symmetry predictions.
 The same colour code, mentioned in the caption of Fig.~\ref{fig1}, is maintained throughout this work. The figures show that the residual symmetry relation predicts a well-defined theoretical correlation between $\theta_{23}$ and $\delta_{\rm CP}$ (yellow region). But in our experimental analysis, corresponding to these predicted regions, we observe the broad band showing the experimentally allowed regions. This broadening is the combined effect of parameter marginalisation as well as the experimental systematics and nuisance parameters used in the analysis through GLoBES. 
%%%%%%%%%%%%%%%%%%%%%%%%%%%%%%%%%%%%%%%%%%%%%%%%%
\begin{figure}[!htbp]
    \centering
    \includegraphics[width=0.49\linewidth]{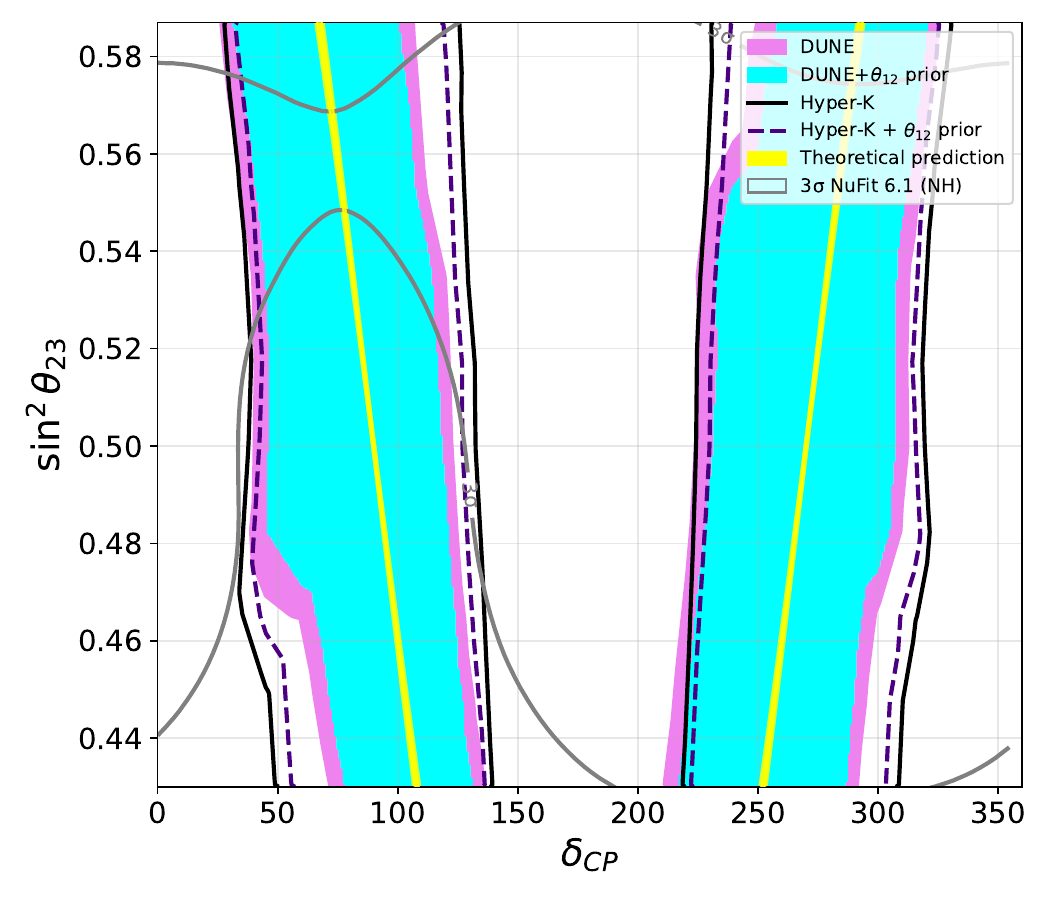}
    \includegraphics[width=0.49\linewidth]{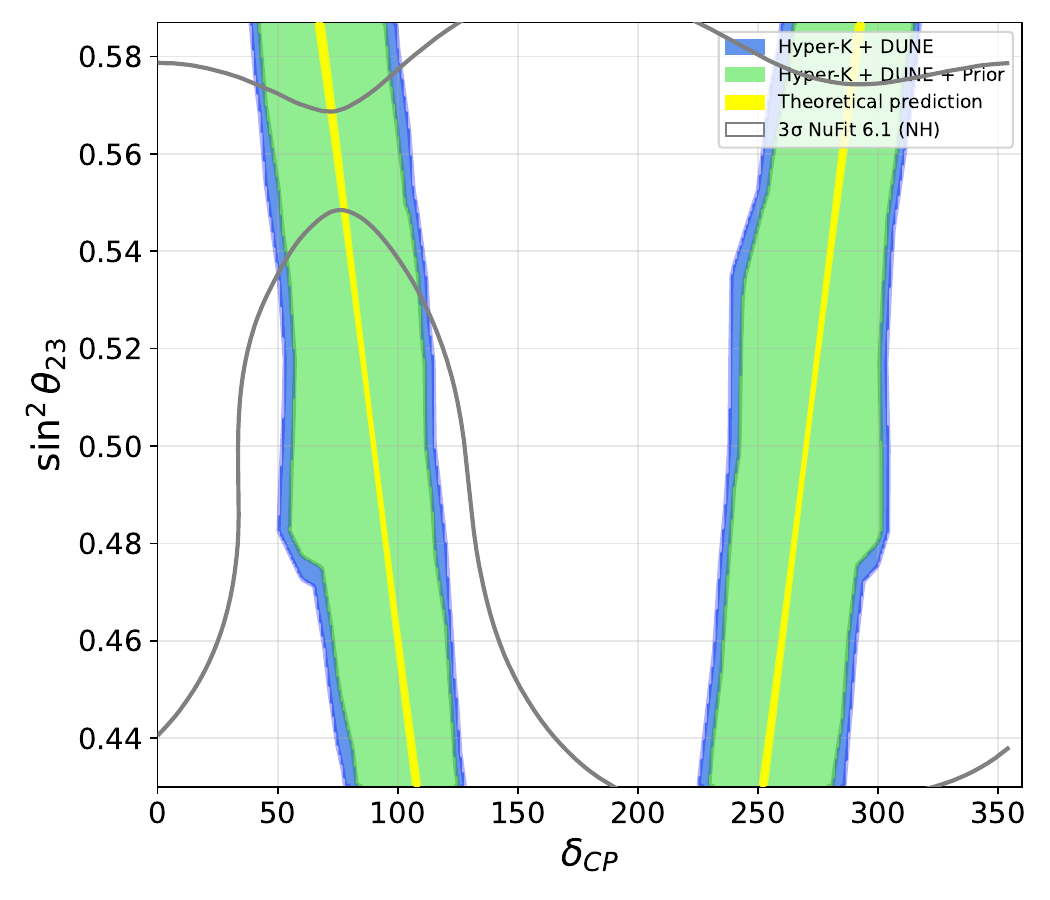}
    \caption{Allowed regions in the $\sin^2\theta_{23}$--$\delta_{\rm CP}$ plane for the fixed-column predictions $C1[1]$ of Table~\ref{tab:1}. The left panels show the sensitivities of DUNE and Hyper-K with and without the $\theta_{12}$ prior, while the right panels display their combined analysis. The yellow band represents the theoretical correlations with the lower and higher $\theta_{13}$ values, as calculated using Eqs.~(\ref{Eqn:7}) and (\ref{Eqn:11}). Shaded regions and black solid contours indicate the expected $3\sigma$ sensitivities. The grey contour lines show the $3\sigma$ allowed region from NuFIT~6.1. The pink and cyan regions correspond to DUNE without and with the $\theta_{12}$ prior, respectively. At the same time, the black solid and indigo dashed contours denote the Hyper-K allowed regions without and with the $\theta_{12}$ prior. In the right panels, the light-blue and light-green regions represent the combined DUNE+Hyper-K sensitivities without and with the $\theta_{12}$ prior, respectively.}
    \label{fig1}
\end{figure}
\begin{figure}[!htbp]
    \centering
\includegraphics[width=0.49\linewidth]{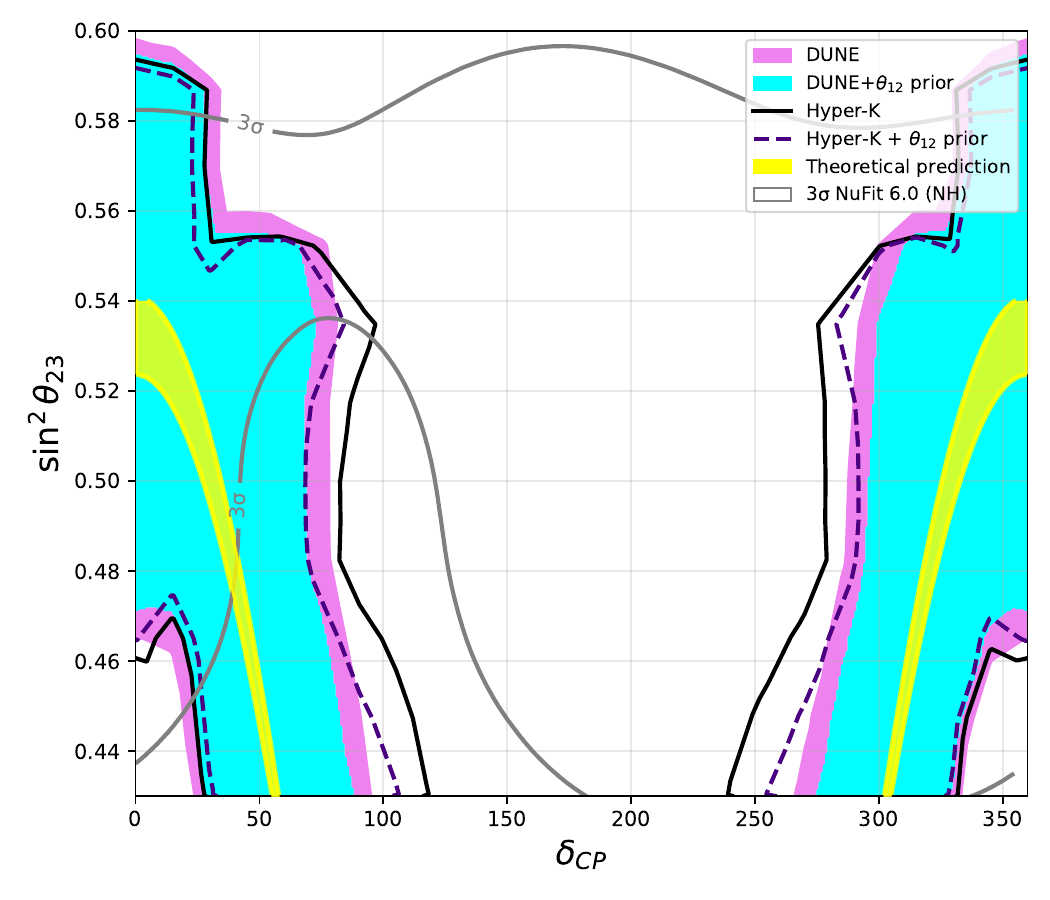}
\includegraphics[width=0.49\linewidth]{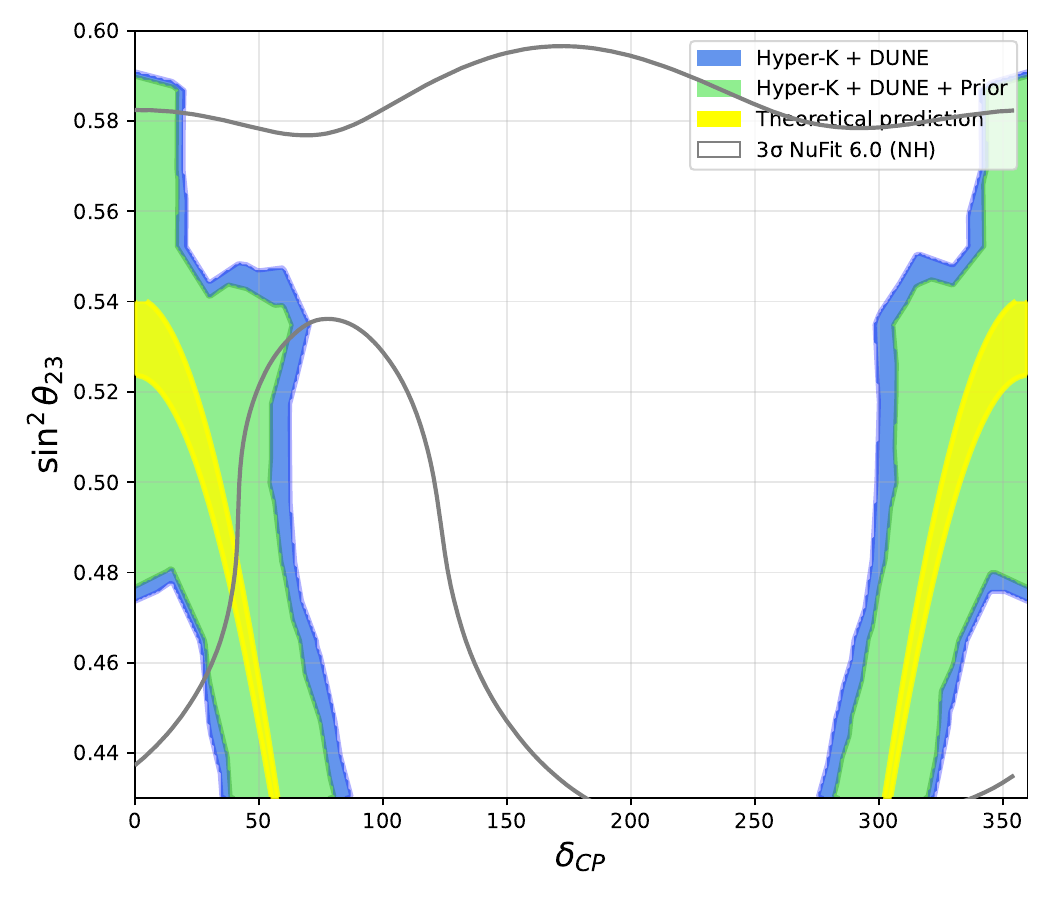}
\includegraphics[width=0.49\linewidth]{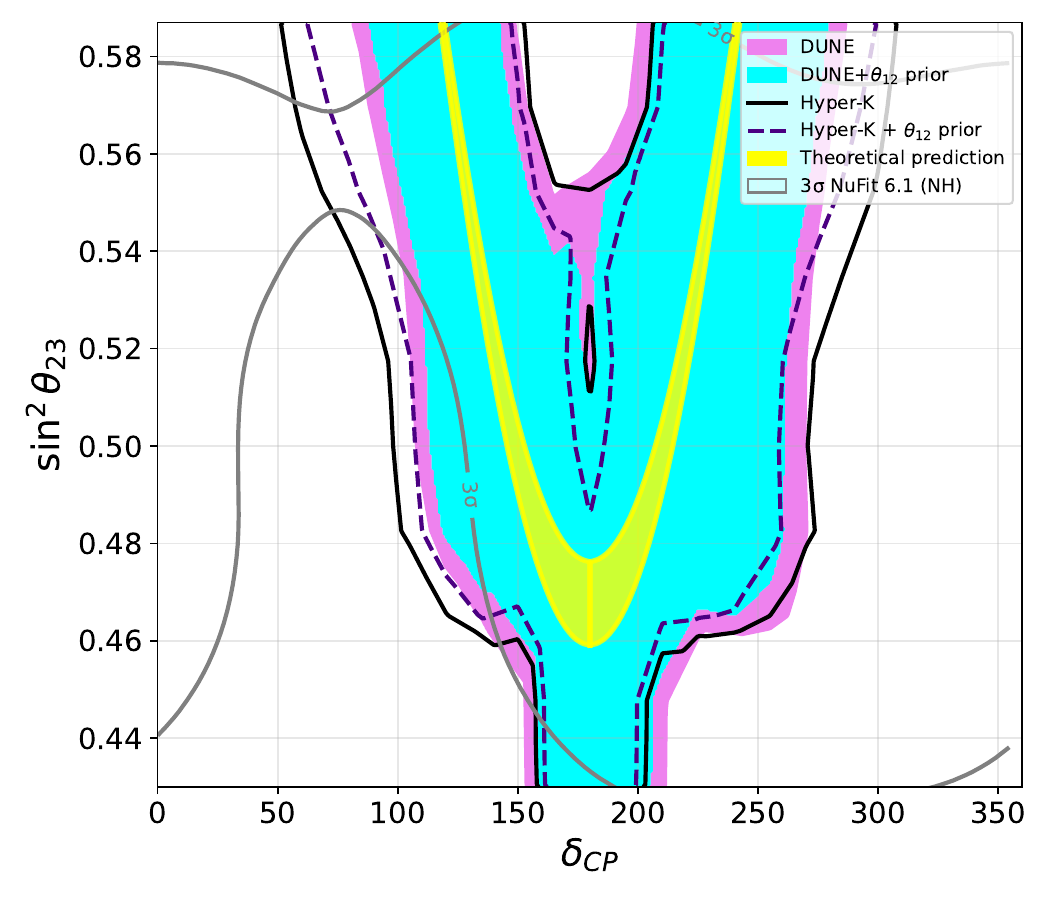}
\includegraphics[width=0.49\linewidth]{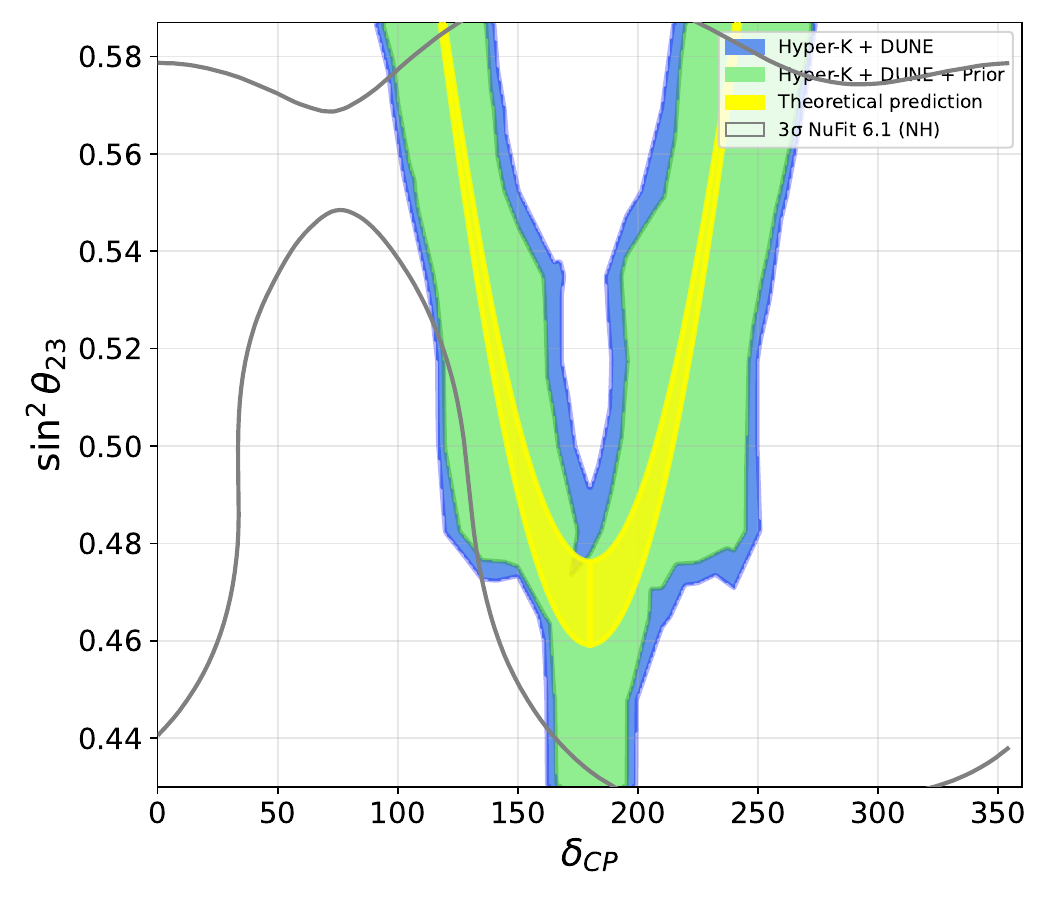}
\caption{Allowed regions in the $\sin^2\theta_{23}$--$\delta_{\rm CP}$ plane for the fixed column predictions $C1[2]$ (upper) and $C1'[2]$ (lower)  from Table~\ref{tab:1}. The experimental sensitivities and theoretical correlations are displayed using the same conventions as in Fig.~\ref{fig1}.}
    \label{fig2}
\end{figure}
%%%%%%%%%%%%%%%%%%%%%%%%%%%%%%%%%%%%%%%%%%%%

Fig.~\ref{fig1} shows the allowed regions in the true $\sin^2\theta_{23}$--$\delta_{\rm CP}$ plane for the fixed-column predictions corresponding to the case $C1[1]$. From the left panels, we find that the relative sensitivities of DUNE and Hyper-K exhibit region-dependent behaviour, with DUNE dominating Hyper-K across the full parameter space. 
%DUNE provides tighter constraints in certain $\delta_{CP}$ intervals, while T2HK achieves comparable or stronger sensitivity in other regions due to its higher event statistics. 
The inclusion of the $\theta_{12}$ prior leads to a modest reduction of the allowed regions for both experiments without qualitatively altering their relative sensitivities. The right panels illustrate the impact of combining the two data sets, where the increased overall statistics significantly reduce the allowed region and improve the determination of $\sin^2\theta_{23}$ and $\delta_{CP}$. With the inclusion of the $\theta_{12}$ prior in the combined DUNE+Hyper-K analysis, the correlated parameter space becomes strongly constrained, with the theoretically expected band lying within the allowed region. In general, the figure reveals that while Hyper-K and DUNE show an increase in statistics for constraints on the parameters individually, their combination leads to a markedly stronger restriction of the allowed parameter space.

\begin{figure}[!htbp]
    \centering
\includegraphics[width=0.49\linewidth]{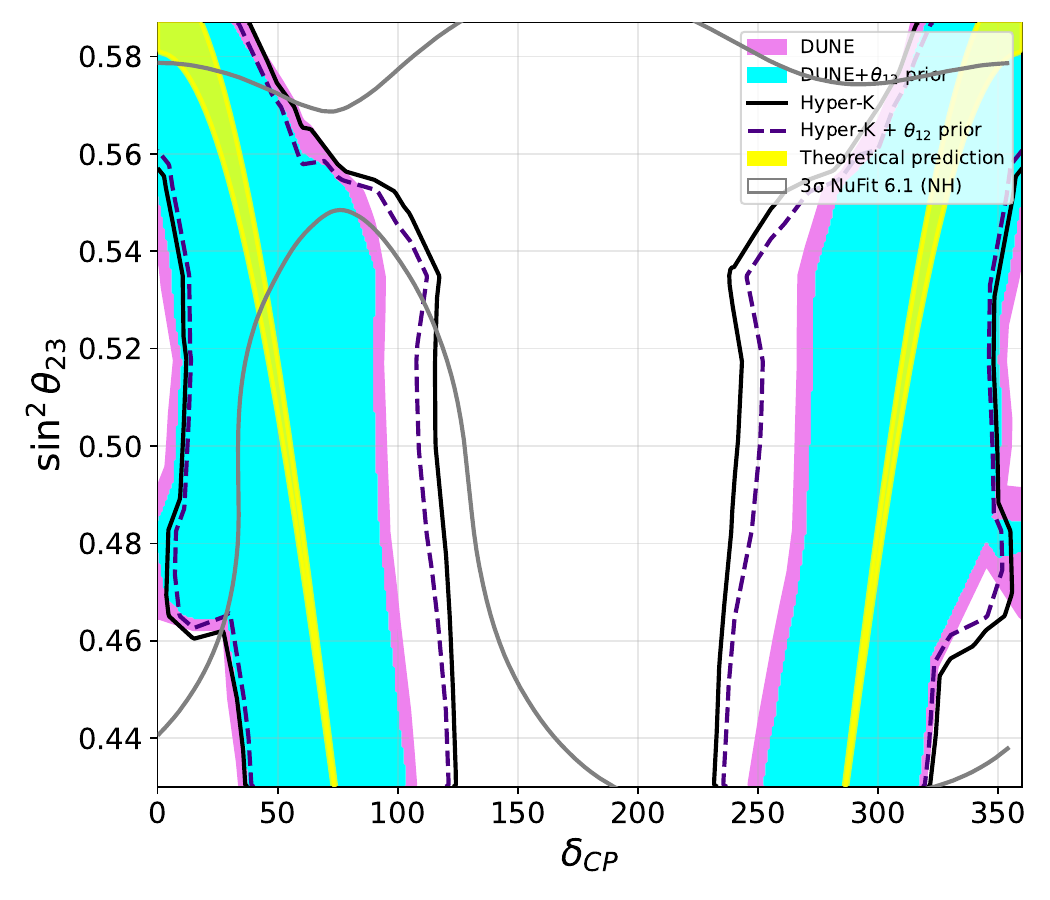}
\includegraphics[width=0.49\linewidth]{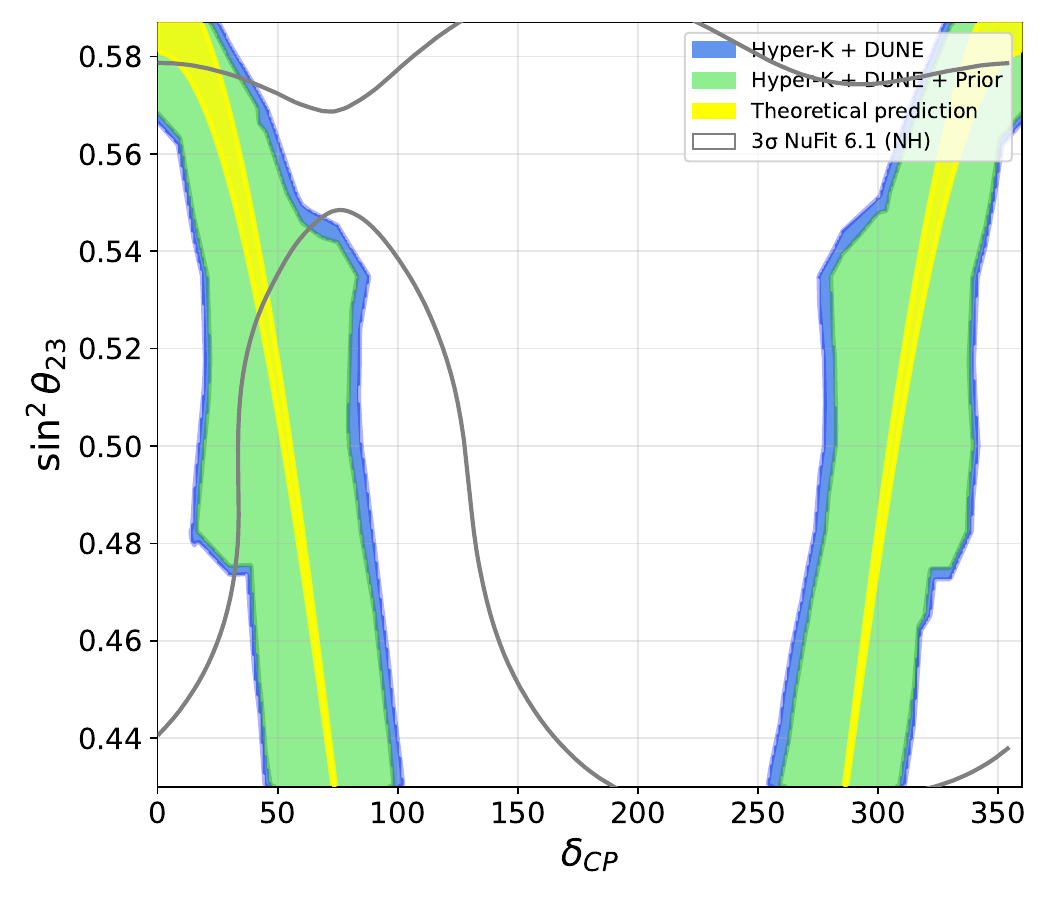}
\includegraphics[width=0.49\linewidth]{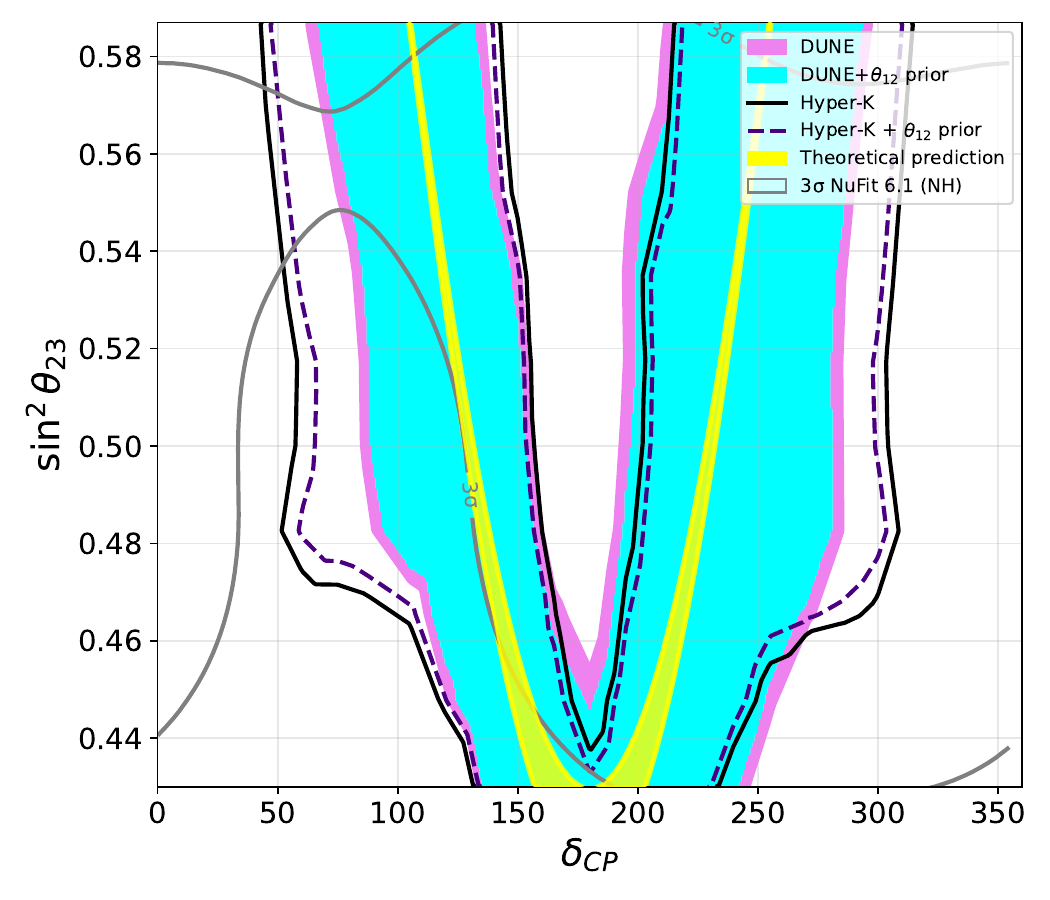}
\includegraphics[width=0.49\linewidth]{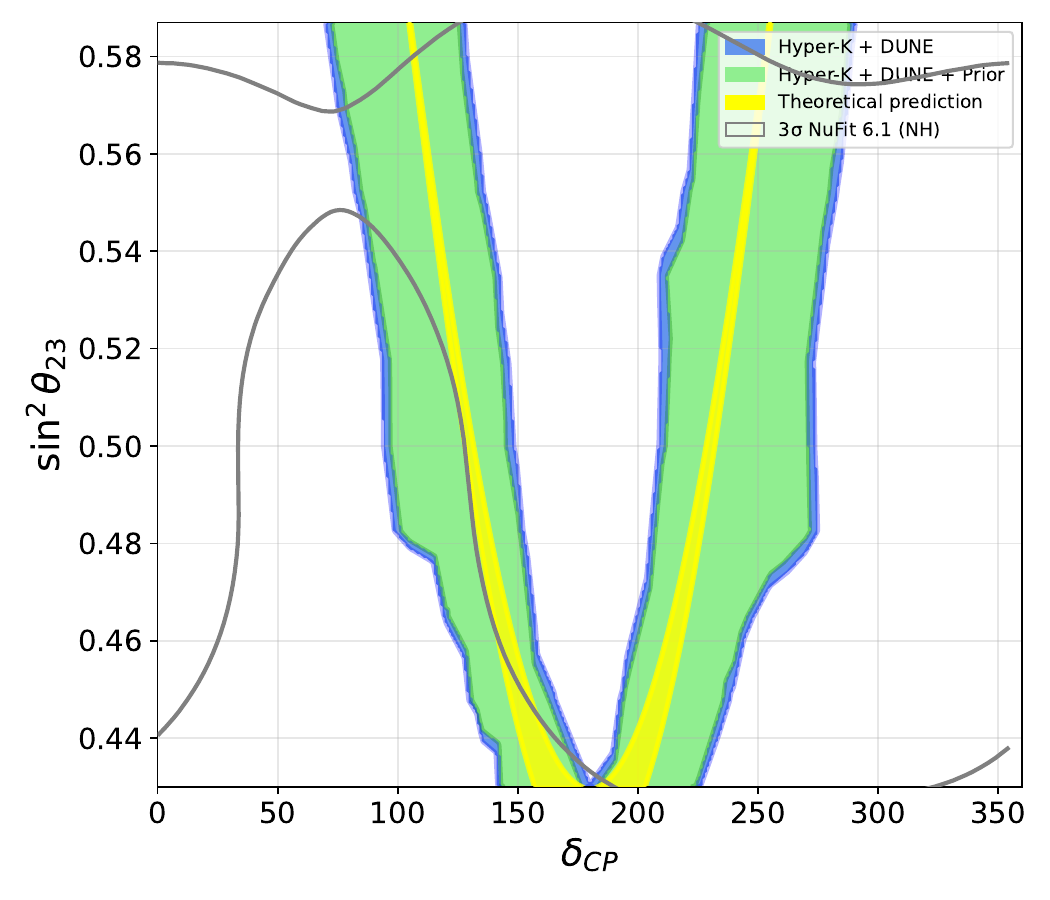}
\caption{Allowed regions in the $\sin^2\theta_{23}$--$\delta_{\rm CP}$ plane for the fixed column predictions $C1[3]$ (upper) and $C1'[3]$ (lower)  from Table~\ref{tab:1}. The experimental sensitivities and theoretical correlations are displayed using the same conventions as in Fig.~\ref{fig1}.}
    \label{fig3}
\end{figure}

\begin{figure}[!htbp]
    \centering
\includegraphics[width=0.49\linewidth]{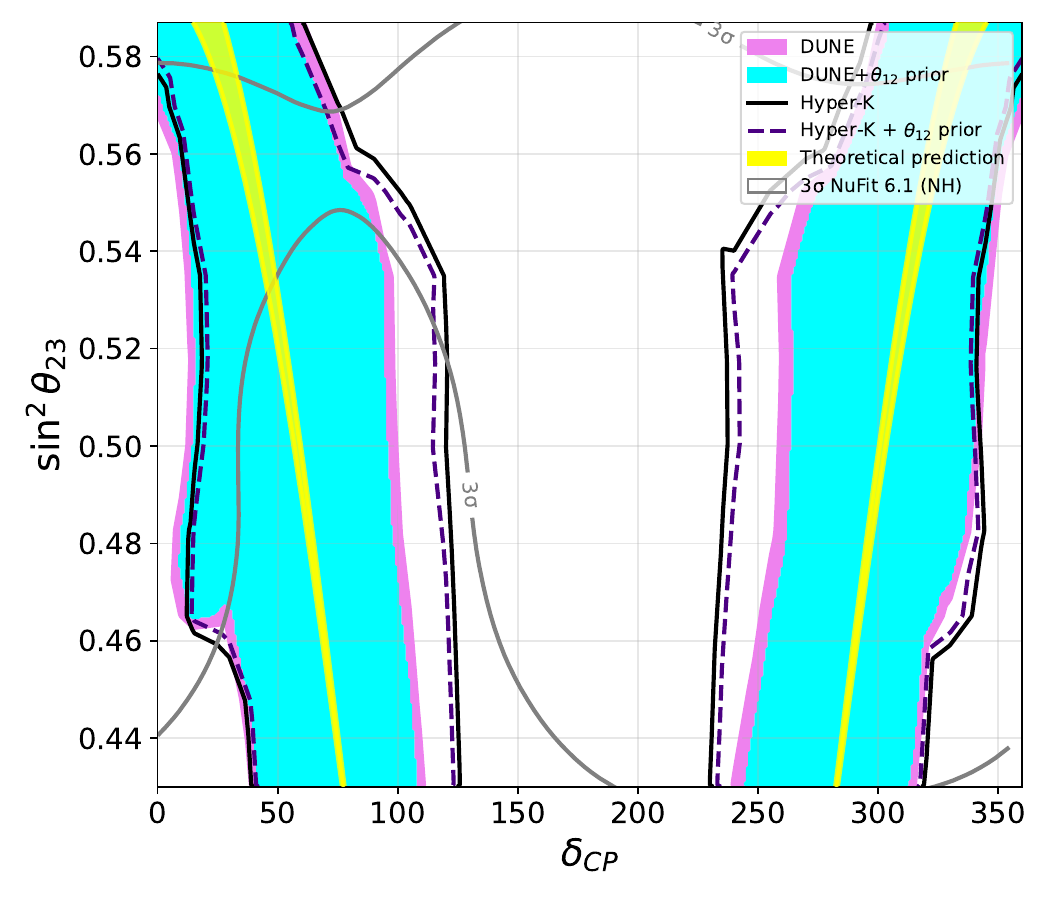}
\includegraphics[width=0.49\linewidth]{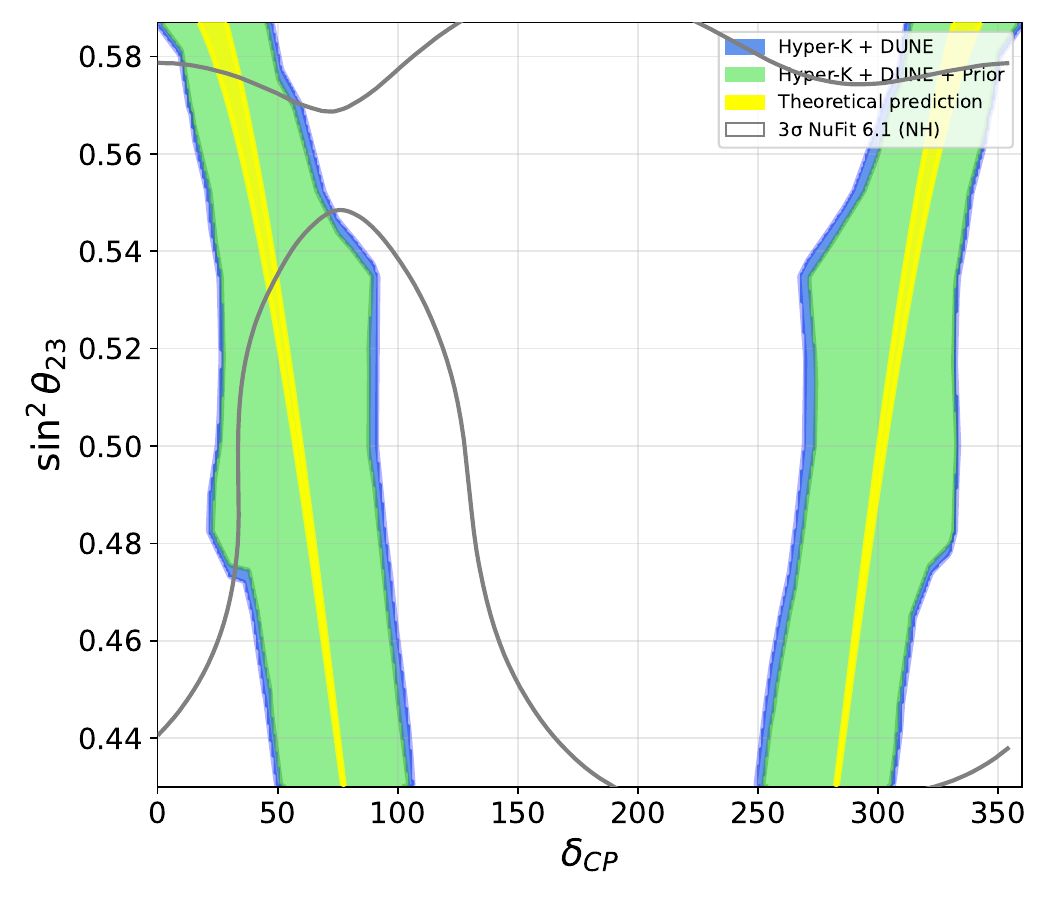}
\includegraphics[width=0.49\linewidth]{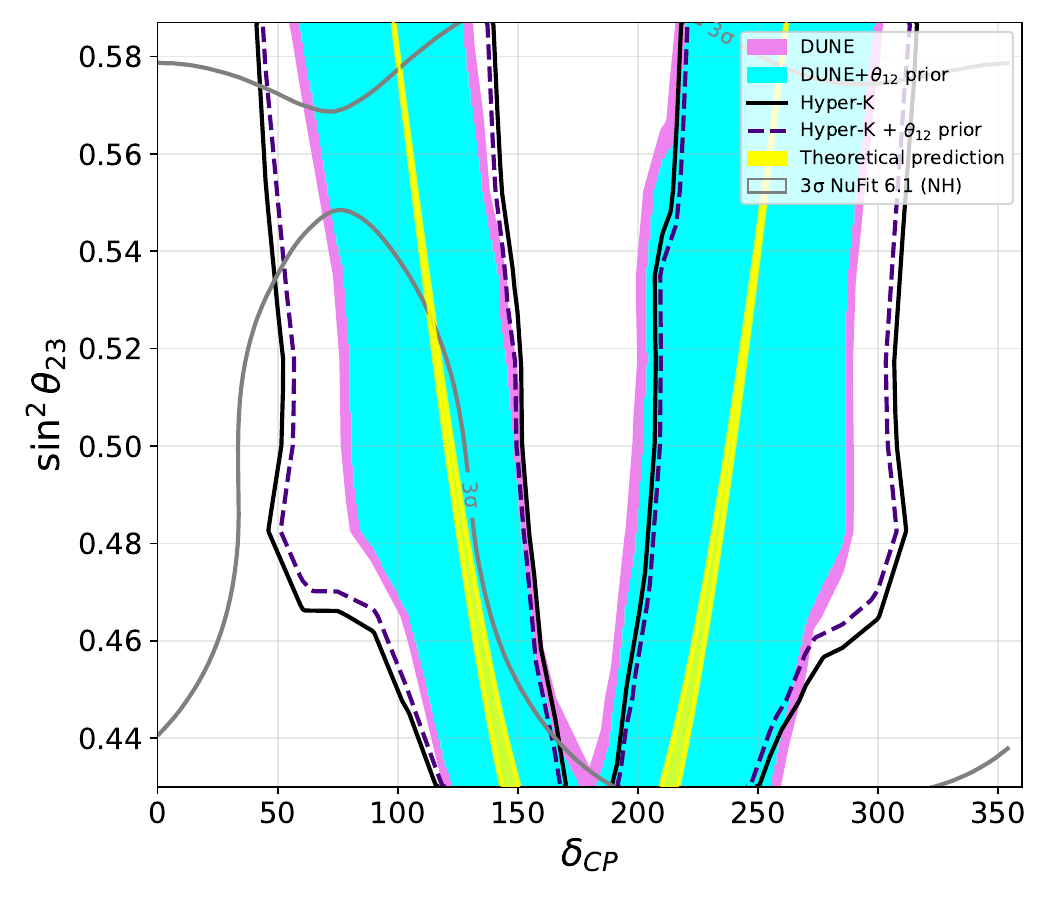}
\includegraphics[width=0.49\linewidth]{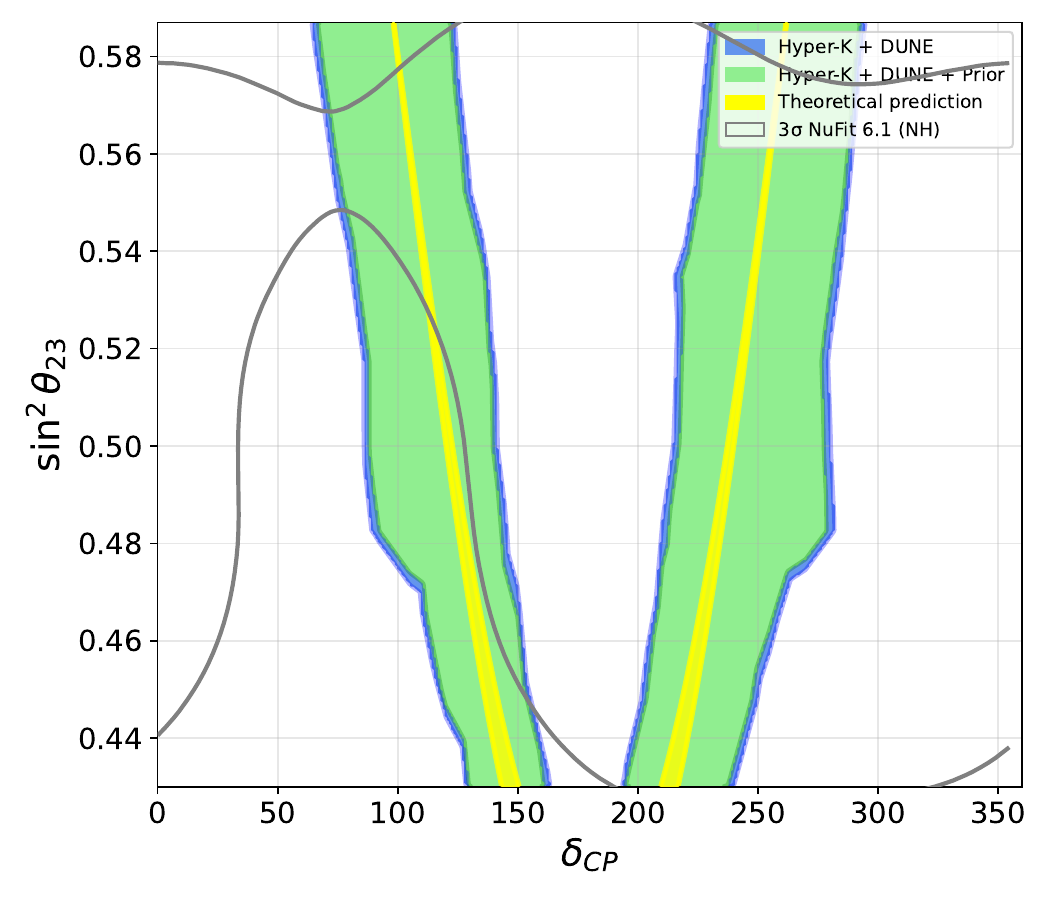}
\caption{Allowed regions in the $\sin^2\theta_{23}$--$\delta_{\rm CP}$ plane for the fixed column predictions $C1[4]$ (upper) and $C1'[4]$( lower)  from Table~\ref{tab:1}. The experimental sensitivities and theoretical correlations are displayed using the same conventions as in Fig.~\ref{fig1}.}
    \label{fig4}
\end{figure}

Fig.~\ref{fig2} depicts the allowed regions for the $C1[2]$ (upper) and $C1'[2]$ (lower) cases. DUNE exhibits better performance and yields a stronger reduction in the allowed parameter space than Hyper-K. The inclusion of the $\sin^2\theta_{12}$ prior visibly enhances the sensitivities of DUNE, Hyper-K, and their combined analysis by further shrinking the allowed regions, highlighting the improved discriminating power arising from the precise determination of $\theta_{12}$. A similar qualitative behaviour is observed in Figs.~\ref{fig3}, \ref{fig4}, and \ref{fig5}. A combination of the two experiments further restricts the allowed parameter space.

%%%%%%%%%%%%%%%%%%%%%%%%%%%%%%%%
\begin{figure}[!htbp]
    \centering
\includegraphics[width=0.49\linewidth]{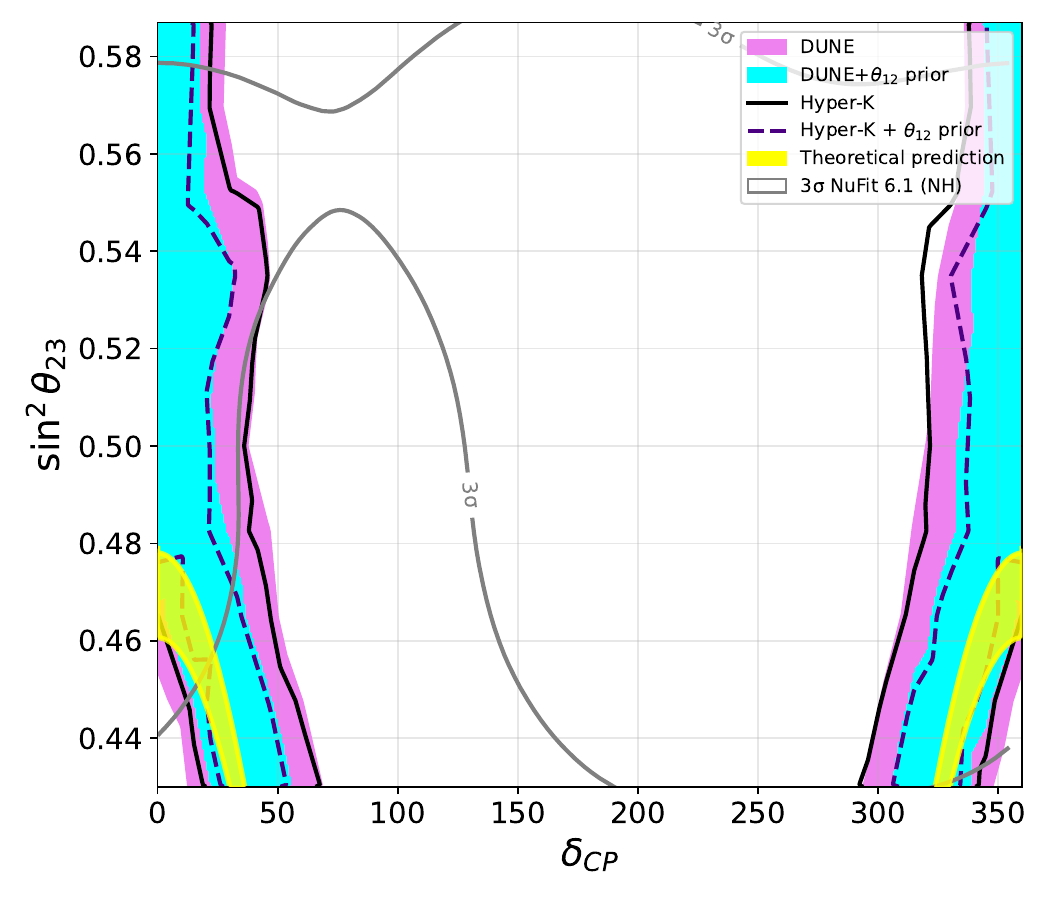}
\includegraphics[width=0.49\linewidth]{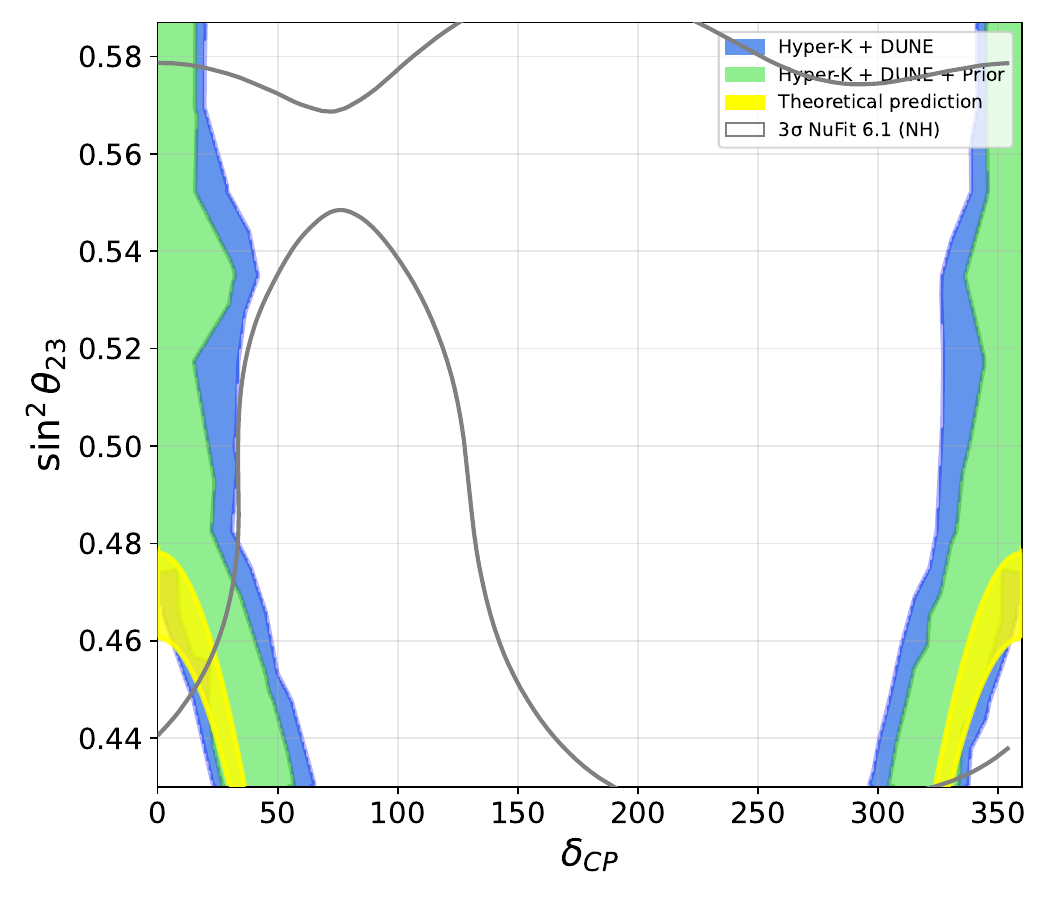}
\includegraphics[width=0.49\linewidth]{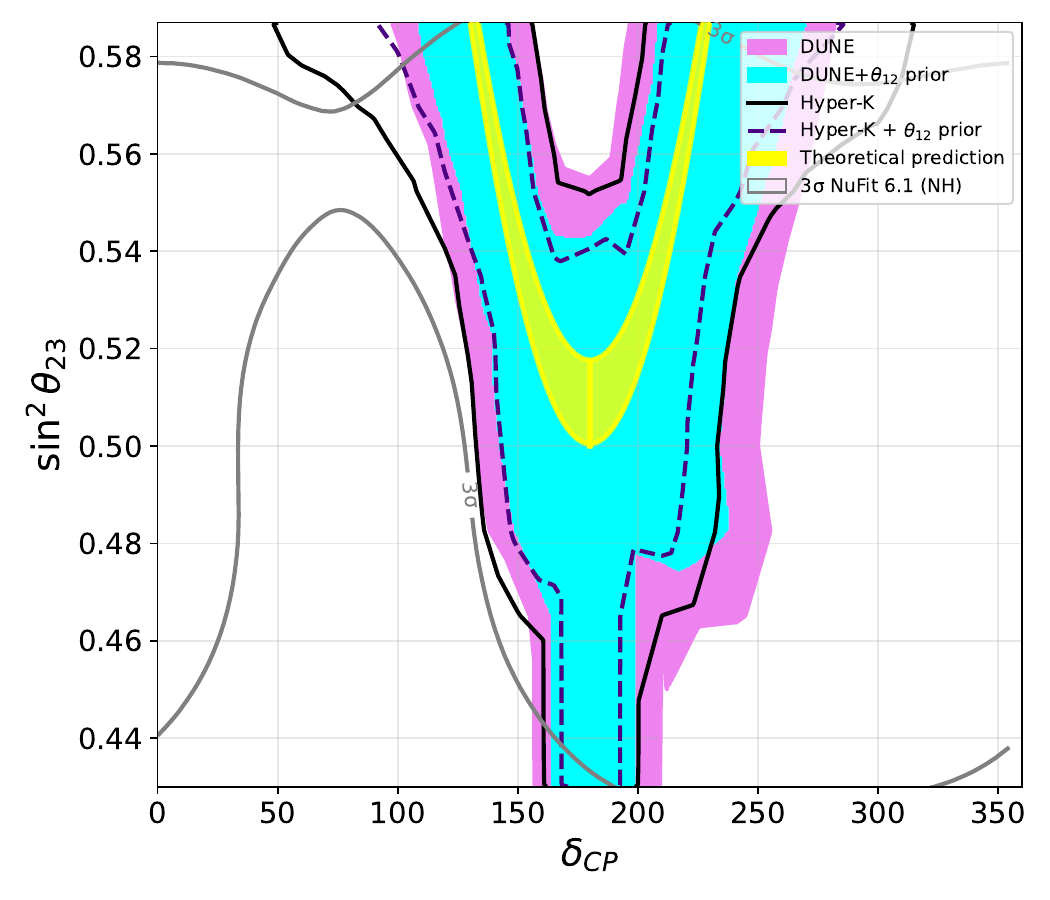}
\includegraphics[width=0.49\linewidth]{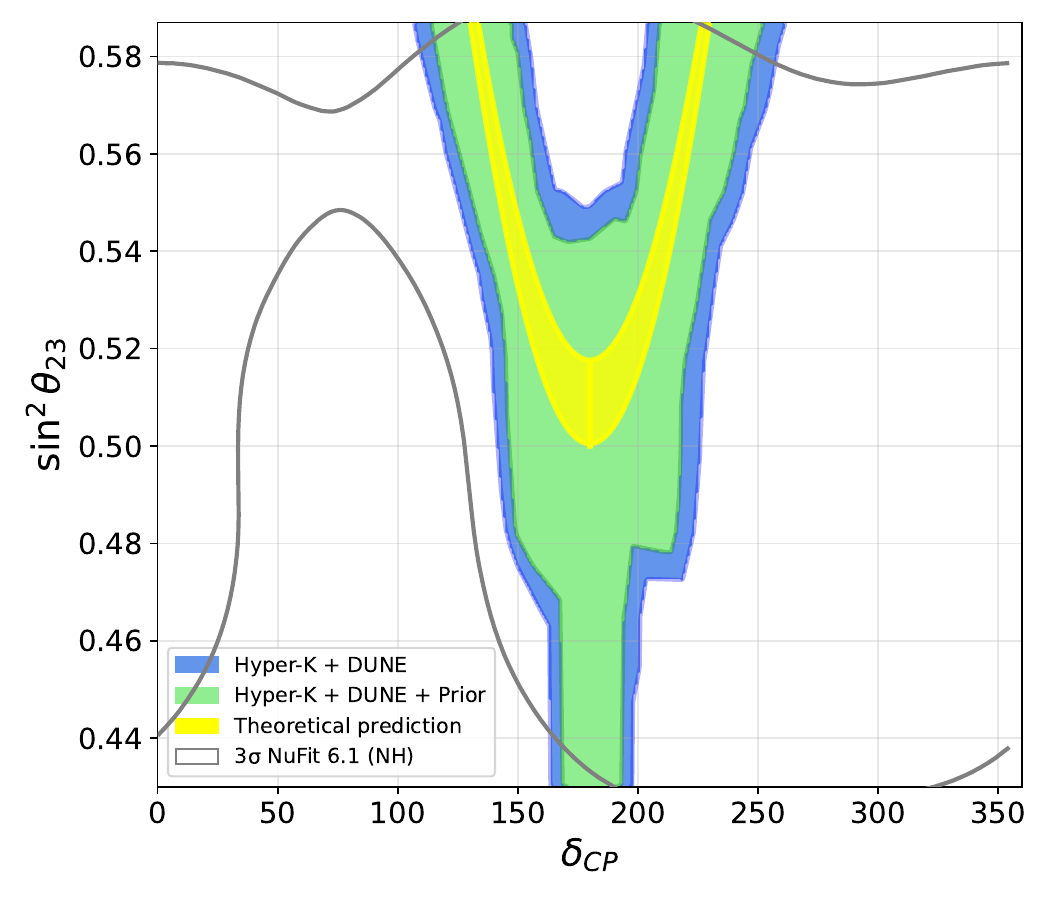}
\caption{Allowed regions in the $\sin^2\theta_{23}$--$\delta_{\rm CP}$ plane for the fixed column prediction $C1[5]$ (upper) and $C1'[5]$ (lower)  from Table~\ref{tab:1}, shown using the same experimental and theoretical conventions as in Fig.~\ref{fig1}.}
\label{fig5}
\end{figure}
%%%%%%%%%%%%%%%%%%%%%%%%%%%%%%%%%%%

A particularly important feature emerges from a comparison of the upper panels of Figs.~\ref{fig2}, \ref{fig3}, \ref{fig4} and \ref{fig5}. For the $C1[2]$ ($C1[3]$) scenario at DUNE, irrespective of the value of $\sin^2\theta_{23}$, true values of the CP phase $\delta_{\rm CP}$ lying approximately in the interval $90^\circ$--$268^\circ$ ($115^\circ$--$245^\circ$) are excluded at the $3\sigma$ level. A similar exclusion range of $115^\circ$--$235^\circ$ is observed for $C1[4]$. In case of $C1[5]$, all true $\delta_{\rm CP}$ in the range $65^\circ$--$290^\circ$ are excluded at 3$\sigma$ irrespective of the true value of $\sin^2\theta_{23}$.
Upon combining DUNE and Hyper-K, these exclusion regions are significantly enlarged. In the combined analysis, for $C1[2]$ ($C1[3]$) and $C1[5]$, true $\delta_{\rm CP}$ values in the approximate ranges $80^\circ$--$280^\circ$ ($100^\circ$--$260^\circ$) and $60^\circ$--$298^\circ$  are excluded at $3\sigma$ for all values of $\sin^2\theta_{23}$. Thus, the strength of the exclusion exhibits a pronounced dependence on $\theta_{23}$. When $\theta_{23}$ lies in the lower octant, the exclusion of true $\delta_{\rm CP}$ values is relatively weak. In contrast, as $\theta_{23}$ moves away from maximal mixing toward the higher end of its allowed range, the exclusion becomes substantially stronger. This qualitative behavior is consistently observed across $C1[2]$, $C1[3]$, $C1[4]$, and $C1[5]$. For Hyper-K, the trend is similar, but the excluded region is smaller than that of DUNE except for $C1[5]$. In the case of $C1[5]$, the performance of both experiments is comparable. 

In contrast, the lower panels of Figs.~\ref{fig2}, \ref{fig3}, \ref{fig4}, and \ref{fig5} exhibit a qualitatively opposite behavior from that observed in the upper panels. When the assumed true value of $\sin^2\theta_{23}$ lies in the lower octant—particularly near the lower edge of the $3\sigma$ allowed range (from NuFit~6.1), DUNE, Hyper-K, and their combination achieve the strongest exclusion of true $\delta_{\rm CP}$ values. The behavior is more prominent in the case of $C1'[2]$ and $C1'[5]$. % In this regime, the exclusion power follows the hierarchy $C1'[5]>C1'[2] > C1'[3] > C1'[4]$.
In the case of $C1'[5]$, Hyper-K exhibits superior performance compared to DUNE within the range $0.44 \leq \sin^2\theta_{23} \leq 0.54$, irrespective of the inclusion of a prior on $\theta_{12}$. However, for $\sin^2\theta_{23} > 0.54$, DUNE outperforms Hyper-K.

From Table~\ref{tab:2}, we further note that although the ranges of $\sin^2\theta_{12}$ allowed by the fixed-column solutions lie within the $3\sigma$ limits of NuFit~6.1, the corresponding best-fit value of $\sin^2\theta_{12}$ often falls outside these theoretically predicted intervals. The greater the separation between the best-fit value and the predicted range, the larger the contribution of the prior term to the total $\chi^2$. Consequently, the inclusion of the JUNO prior on $\sin^2\theta_{12}$ shifts the minimum $\chi^2$ to higher values. In some of the scenarios, such as $C1[6]$ and $C1'[6]$ (shown in the appendix), this additional contribution drives the total $\chi^2_{\min}$ beyond the $3\sigma$ threshold ($\chi^2_{\min} > 11.83$), resulting in the complete exclusion of the corresponding parameter space when the JUNO prior is added. The allowed regions are also significantly reduced. This demonstrates that, in the presence of precise measurements of $\sin^2\theta_{12}$, the combined experimental setup can exclude a subset of fixed-column predictions. Furthermore, improved precision from future JUNO data is expected to strengthen these exclusions and place tighter constraints on a broader class of theoretical scenarios.

%In Fig.~\ref{fig5}, we have presented the results for $C1[7]$ and $C1[8]$ and observe that most of the true $\sin^2\theta_{23}$--$\delta_{\rm CP}$ plane is excluded in both the cases. Both DUNE and T2HK exhibit similar exclusion behaviour, and the allowed region appears around the CP-conserving values, irrespective of the value of $\sin^2\theta_{23}$. The band further shrinks if DUNE is combined with T2HK. If prior is added, then all the parameter space gets excluded in case of $C1[7]$, while in case of $C1[8]$, the allowed band almost shrinks to a line appearing at $\delta_{\rm CP} = 0^\circ$ or $360^\circ$.

%This indicates that, when combined, the experimental sensitivities are especially effective in excluding large portions of the CP phase, thereby considerably enhancing the discriminating power.

\subsection{$\delta_{\rm CP}$ exclusion reach of DUNE and Hyper-K}
The high-precision measurements expected from upcoming neutrino experiments provide a powerful opportunity to confront theoretical models with data. We assume that by the time DUNE and Hyper-K complete their data taking, the atmospheric mixing angle $\theta_{23}$ will be precisely known. In this context, we address the following questions: Can DUNE or Hyper-K rule out a given theoretical model at a confidence level exceeding $3\sigma$ for some fraction of the true leptonic CP phase $\delta_{\rm CP}$? If so, how does this exclusion fraction evolve with the experimental exposure? To answer these questions, we define the exclusion fraction of $\delta_{\rm CP}$ for a fixed true value of $\theta_{23}$ as follows:
\begin{equation}
f_{\rm exclude}(\text{exposure}) \equiv 
\frac{N_{\delta_{\rm CP}}\!\left(\chi^2_{\min}(\delta_{\rm CP}) > \chi^2_{3\sigma}\right)}
{N_{\delta_{\rm CP}}^{\rm total}},
\end{equation}
where, the numerator, $N_{\delta_{\rm CP}}(\chi^2_{\min}>\chi^2_{3\sigma})$, is the number of those values of true $\delta_{\rm CP}$ for which the minimized $\chi^2_{min}$ is greater than 3$\sigma$ and $N_{\delta_{\rm CP}}^{\rm total}$ in the denominator is the number of \textit{tested true values of $\delta_{\rm CP}$}. 
For a given set of true values of the oscillation parameters, the $\chi^2_{min}$ is obtained after full marginalisation over the allowed model parameters as discussed in section~\ref{chi2}, optimising over the neutrino-antineutrino runtime. The term \textit{"tested true values of $\delta_{\rm CP}$"} refers to the discrete set of values of $\delta_{\rm CP}$ for which we simulate the expected DUNE and Hyper-K data and test the compatibility of each theoretical model. The total exposure of DUNE is assumed to be 13 years and is equally divided between neutrino and antineutrino running \cite{DUNE:2020jqi}. For Hyper-K, we consider a total exposure of 10 years, with the neutrino-to-antineutrino runtime ratio fixed at 1:3 \cite{Hyper-Kamiokande:2018ofw}.

%%%%%%%%%%%%%%%%%%%%%%%%%%%%%%%%%%%%%%%%%%%%%%%%
\begin{figure}[!htbp]
    \centering
\includegraphics[width=1.02\linewidth]{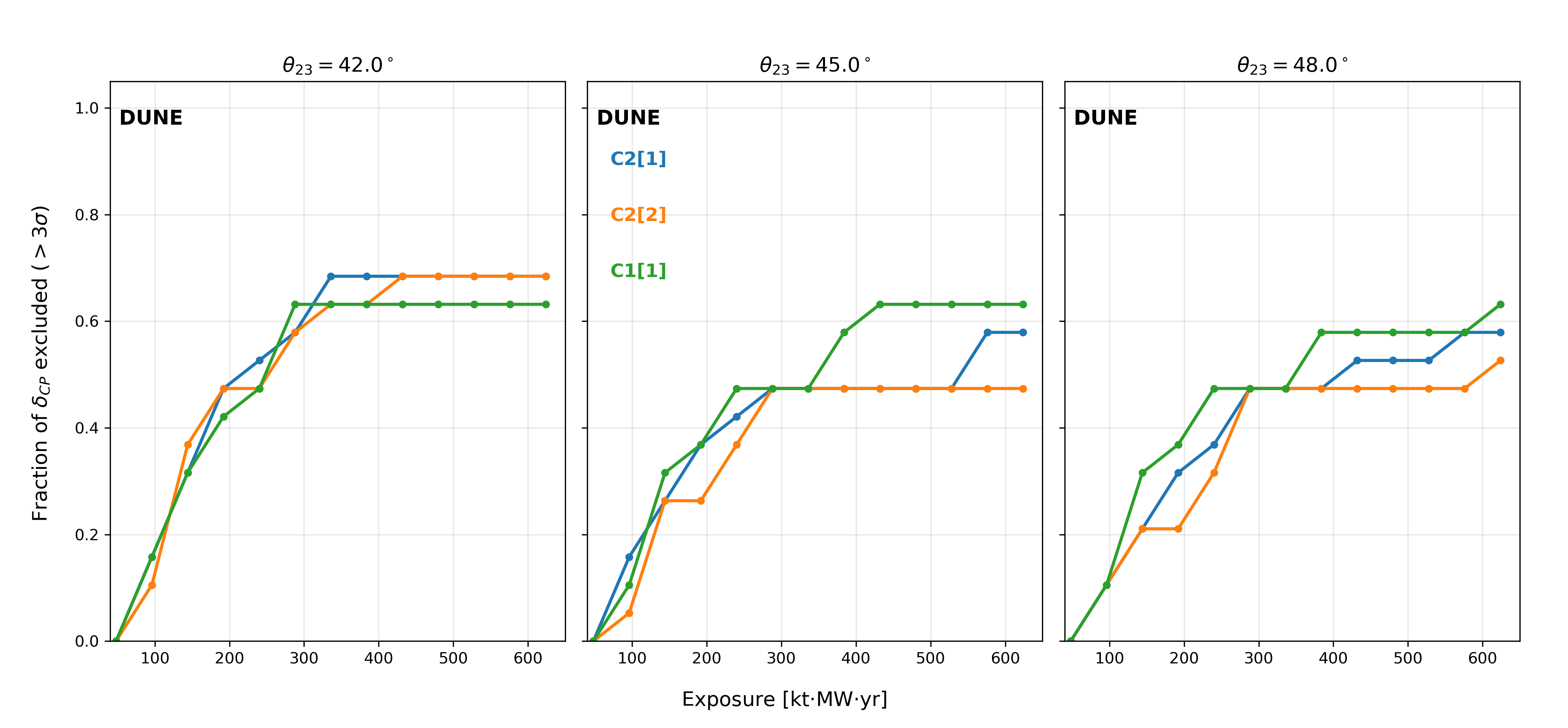}
\includegraphics[width=1.02\linewidth]{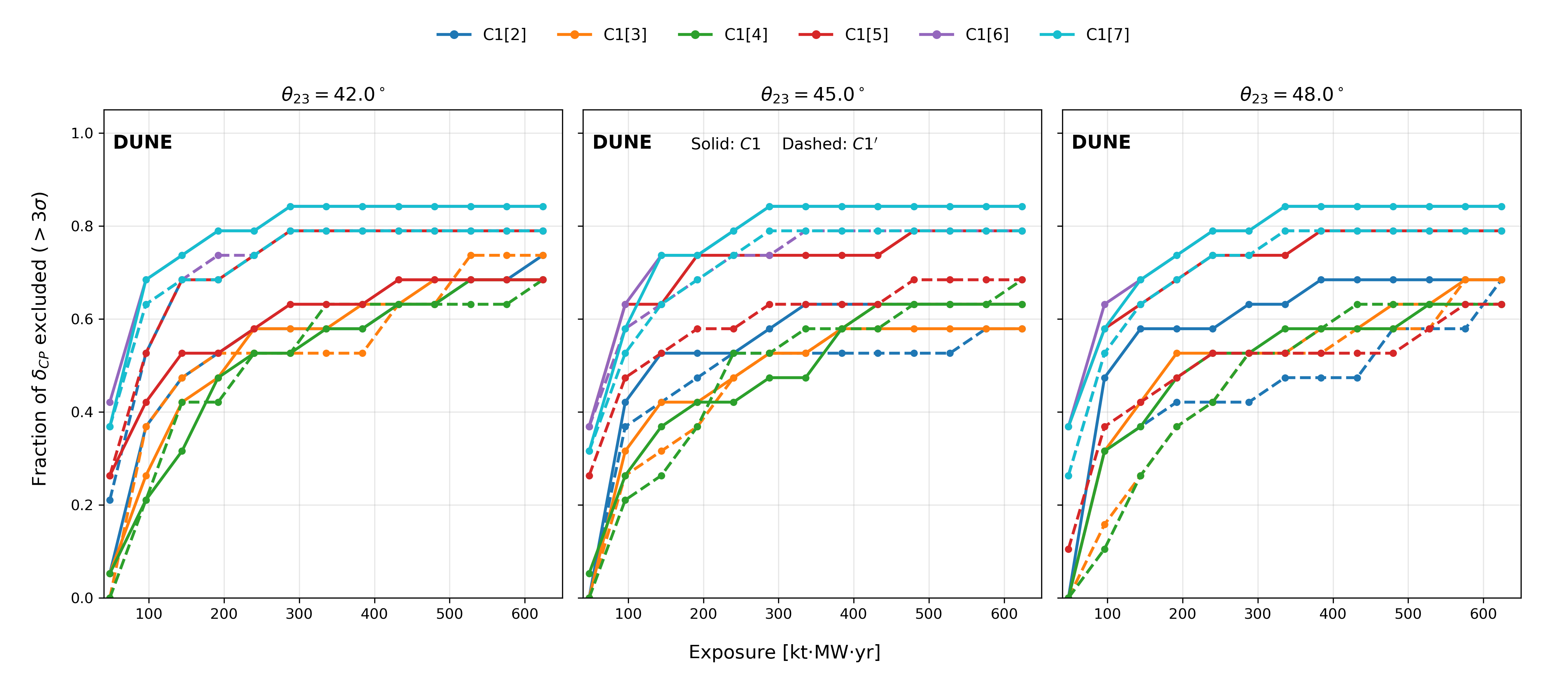}
\caption{Fraction of the true leptonic CP phase $\delta_{\rm CP}$ excluded at more than $3\sigma$ as a function of the DUNE exposure for the models $C1[2]$--$C1[7]$ ($C2[1]$, $C2[2]$, and $C1[1]$) shown in the second row (first row). The three panels in each row correspond to $\theta_{23}=42.0^\circ$ (left), $45.0^\circ$ (middle), and $48.0^\circ$ (right). Solid (dashed) curves denote the $C1$ ($C1'$) predictions in the second row. The total exposure corresponds to 13 years of operation, equally divided between neutrino and antineutrino running. The colour coding of the predictions is indicated in the middle panel of each row.}
    \label{fig6}
\end{figure}
\begin{figure}[!htbp]
    \centering
\includegraphics[width=1.02\linewidth]{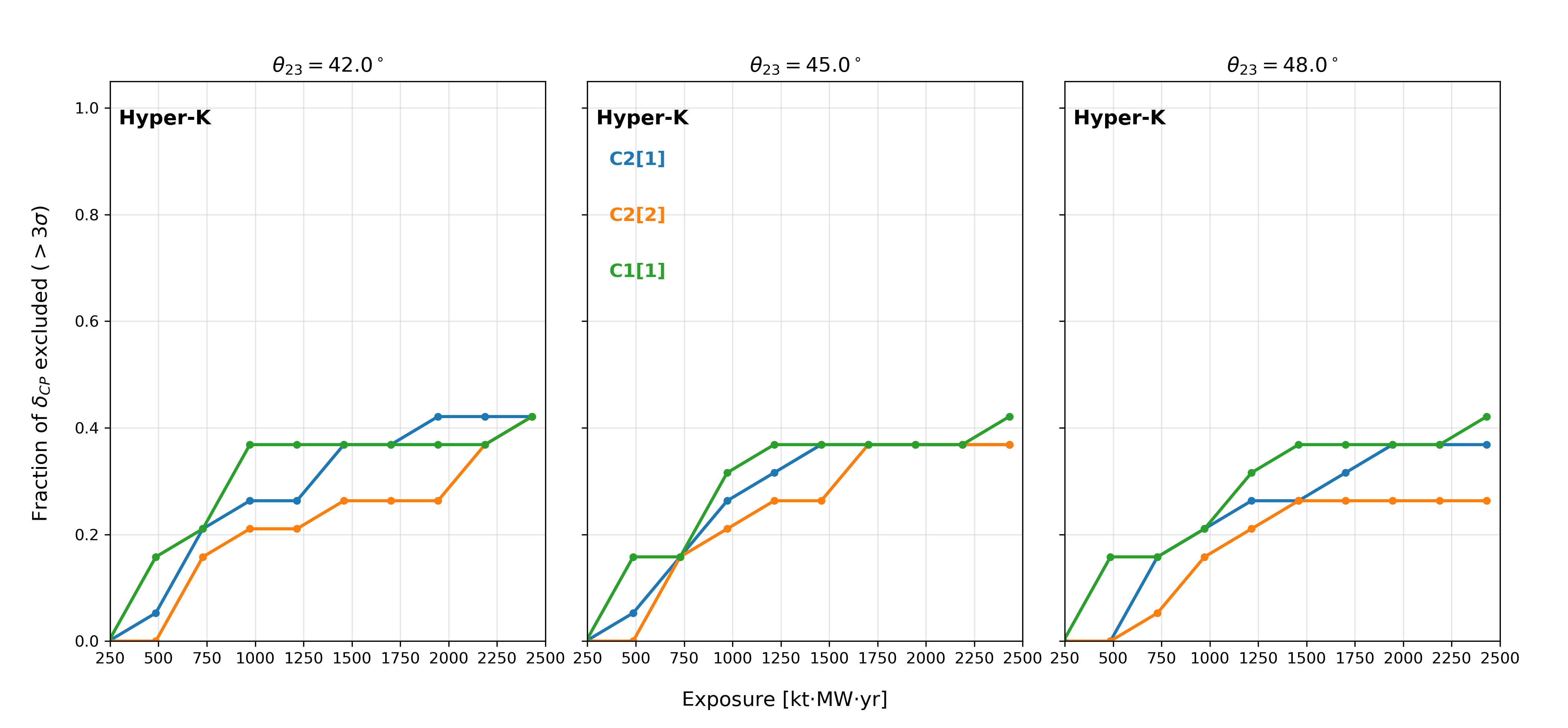}
\includegraphics[width=1.02\linewidth]{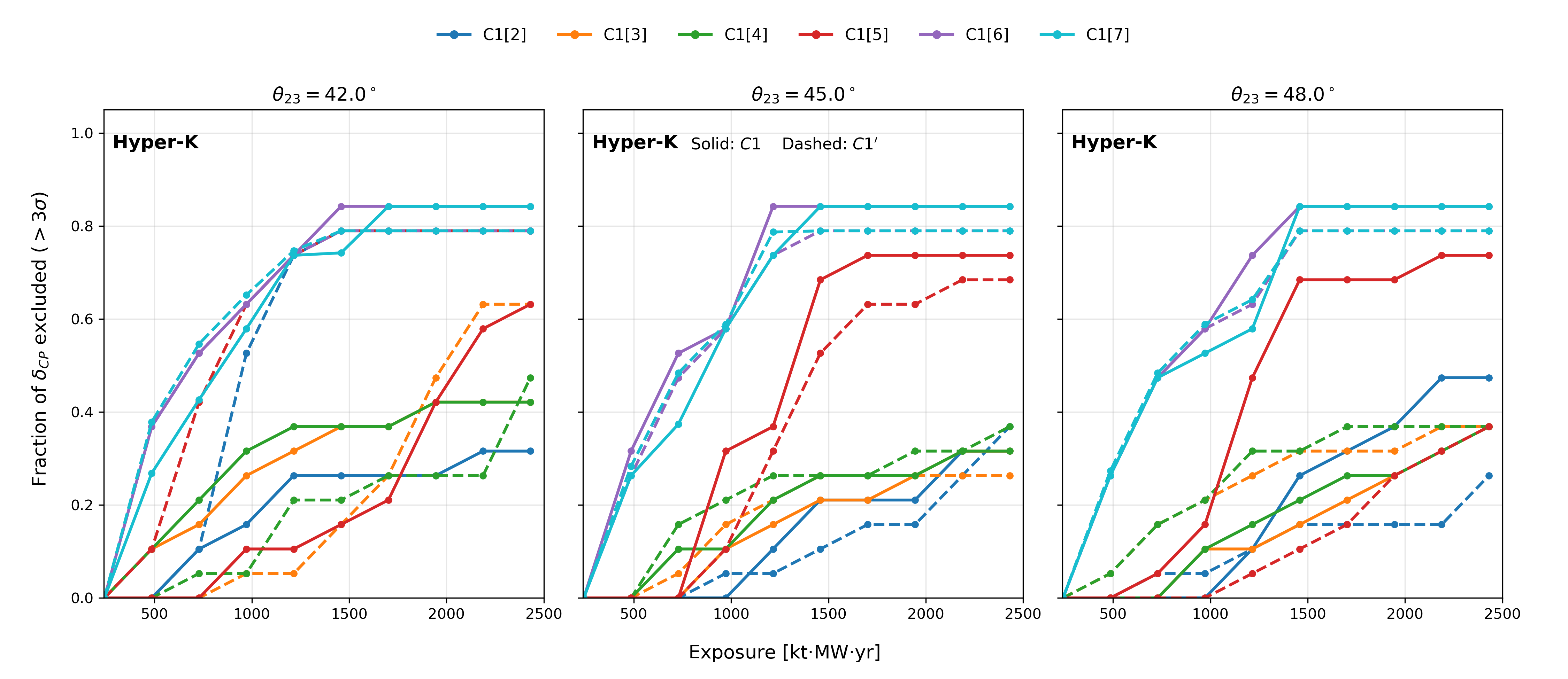}
\caption{Fraction of the true leptonic CP phase $\delta_{\rm CP}$ excluded at more than $3\sigma$ as a function of the Hyper-k exposure. The total exposure corresponds to $10$ years of operation, divided between neutrino and antineutrino modes in a $1{:}3$ ratio. The color coding of the predictions is indicated in the middle panel of each row.}
    \label{fig7}
\end{figure}
%%%%%%%%%%%%%%%%%%%%%%%%%%%%%%%%%%%%%%%%%%%%%%%%%%%
This observable $f_{\rm exclude}$ helps to identify the potential of these future generation experiments. It directly measures the degree to which DUNE/Hyper-K can falsify a given theoretical framework. It also indicates the strength of the experiments in terms of discrimination power as well as model predictivity. The evolution of this quantity with exposure provides a quantitative assessment of the exclusion potential of DUNE and Hyper-K for each model under consideration.

In Fig~\ref{fig6} and \ref{fig7}, assuming three representative choices of the atmospheric mixing angle, $\theta_{23}=42.0^\circ$, $45.0^\circ$, and $48.0^\circ$, it can be seen that DUNE and Hyper-K can exclude the fraction of true $\delta_{\rm CP}$ at more than $3\sigma$ as a function of exposure. Further, we quantify the relative strength of exclusion by the rate at which this fraction grows with exposure, such that for a given exposure, larger values of $f_{\rm excl}$ indicate stronger exclusion. Although this behavior is entirely model dependent, in most cases the true value $\theta_{23}=42.0^\circ$ in the lower octant exhibits a stronger exclusion than its higher-octant counterpart, $\theta_{23}=48.0^\circ$. 

The smaller allowed region, and hence the strongest exclusion behavior, can be directly traced back to the structure of the allowed regions in the $\sin^2\theta_{23}$–$\delta_{\rm CP}$ plane. In particular, the very small parameter regions predicted by the $C1[6]$ and $C1[7]$ models (see Figs.~\ref{fig8} and \ref{fig9}) lead to an early and rapid loss of compatibility with experimental data as the exposure of upcoming experiments increases. Consequently, the fraction of true $\delta_{\rm CP}$ values that can be excluded at $3\sigma$ grows quickly with exposure and saturates at high values. The dependence on $\theta_{23}$ follows from the same geometric origin: when $\theta_{23}$ lies in the lower octant, the tension between the model predictions and the experimental sensitivities is strongest, leading to the fastest rise in the exclusion fraction, while for larger $\theta_{23}$ the growth is more gradual before reaching comparable saturation. 

The $\delta_{\rm CP}$ exclusion behavior discussed earlier for the $C1[2]$ and $C1'[2]$ (and similarly for  $C1[5]$ and  $C1'[5]$ ) cases is also borne out by this analysis. It can be seen in Figs.~\ref{fig6} and \ref{fig7} that the primed scenario attains a higher exclusion fraction than the unprimed case. In contrast, when $\theta_{23}$ is above or equal to $45.0^\circ$, the unprimed cases exhibit stronger exclusion and rise somewhat faster than the primed cases.
This qualitative behavior is consistent for both DUNE and Hyper-K. Moreover, the growth of the exclusion fraction $f_{\rm exclude}$ is significantly steeper at early exposures for $\theta_{23}=42.0^\circ$ compared to its behavior at $\theta_{23}=45.0^\circ$ and $\theta_{23}=48.0^\circ$.

%The predictions C2[1] and C2[2] display weak behaviour and as observed, for $\theta_{23}=41.5^\circ$ they gradually become testable and reach exclusion fractions of $\sim 65$--$75\%$ at large exposure. Their exclusion ability substantially degraded for $\theta_{23}=48.5^\circ$.\\

From this analysis, we observe that DUNE exhibits a stronger exclusion capability in most cases. Fixing the atmospheric mixing angle $\theta_{23}$ significantly enhances the exclusion capability of these experiments, as one of the primary objectives of these two experiments is the discovery of leptonic CP violation. The intrinsic degeneracy between $\theta_{23}$ and the CP phase $\delta_{\rm CP}$ dilutes CP sensitivity; therefore, assuming a precise determination of $\theta_{23}$ leads to a substantial sharpening of the CP measurement and, consequently, to a markedly higher exclusion fraction. Overall, the observed model-dependent behavior demonstrates that future long-baseline experiments such as DUNE and Hyper-K will achieve substantial predictive power, enabling precise tests of neutrino frameworks over a wide region of parameter space.

\section{Summary and Discussion}\label{sec:summary}
 We have investigated the fixed-column predictions for the leptonic mixing matrix arising from residual symmetries from discrete subgroups of $\mathrm{U}(3)$ and modular symmetries that are consistent with current neutrino oscillation data. Such constructions lead to well-defined correlations among the leptonic mixing parameters, most notably between the atmospheric mixing angle $\theta_{23}$ and the Dirac CP phase $\delta_{\rm CP}$, providing sharp correlations for experimental tests. Incorporating the recent JUNO measurement of $\sin^{2}\theta_{12}$, we showed that a number of fixed-column predictions are already disfavored, while several fixed-column scenarios remain viable as per NuFIT 6.1 global analysis results.

For all these cases, we performed a detailed sensitivity analysis using prospective data from the long-baseline experiments DUNE and Hyper-K, both individually and in combination. The allowed regions in the $\sin^{2}\theta_{23}$–$\delta_{\rm CP}$ plane, shown in Figs.~\ref{fig1}–\ref{fig5} (also see Fig.~\ref{fig8}, \ref{fig9}, and \ref{fig10} in Appendix~\ref{app:plots}), illustrate how the theoretical correlations are broadened by experimental effects and parameter marginalization, and how the inclusion of the JUNO prior significantly contracts—or in some cases fully excludes—the experimentally allowed parameter space. The combined DUNE+Hyper-K setup consistently yields the strongest constraints, underscoring the complementarity of the two experiments. To further quantify the testing power of future facilities, we calculated the $\delta_{\rm CP}$ exclusion fraction and studied its evolution with experimental exposure. As shown in Fig.~\ref{fig6}, DUNE (and Hyper-K in Fig.~\ref{fig7}) can exclude a large fraction of the CP phase space at the $3\sigma$ level for many fixed-column predictions, with the exclusion capability strongly dependent on the true value of $\theta_{23}$. In particular, scenarios exhibiting strong restrictions in the $\theta_{23}$–$\delta_{\rm CP}$ plane correspond to rapidly increasing exclusion fractions, reaching $\mathcal{O}(80\text{--}90\%)$ for realistic exposures.

Our results demonstrate that the precise determination of $\theta_{12}$ can significantly reduce the allowed parameter space and can disfavour some of these theoretical predictions. The combined precision of JUNO, DUNE, and Hyper-K provides a powerful and comprehensive framework for scrutinising residual-symmetry-based flavour models. The interplay of precise $\theta_{12}$ measurements by JUNO and the sensitivity of long-baseline accelerator neutrino experiments to $\theta_{23}$ and $\delta_{\rm CP}$ is expected to decisively test—and potentially exclude—a wide class of discrete-symmetry predictions for leptonic mixing in the near future.

\section{Acknowledgement}
 This work is supported by the Department of Space (DOS), Government of India. SG acknowledges the J.C. Bose Fellowship (JCB/2020/000011) of the Science and Engineering Research Board of the Department of Science and Technology, Government of India, and the Department of Space, Govt. of India. DD thanks the three National Science Academies of India for the Focus Area Science Technology Summer Fellowship 2025, which enabled this work to be carried out at the Physical Research Laboratory, Ahmedabad.  MK would like to acknowledge Physical Research Laboratory, Ahmedabad, for the Post-Doctoral Fellowship. The numerical computations and simulations were performed on the Param Vikram-1000 High Performance Computing Cluster of the Physical Research Laboratory (PRL).

\appendix
\section{Fixing a column of $U_{\rm PMNS}$ from residual symmetries} \label{app:DFS_background}
We outline in this Appendix the group-theoretical method for predicting one or more elements of the leptonic mixing matrix, $U_{\rm PMNS}$. The three generations of neutrinos and charged leptons are massive in the $SU(3)_C \times U(1)_Q$ phase of the theory. The theory is assumed to possess an additional global discrete symmetry under which the three generations of the left-chiral charged leptons, $e_{Li}$ and neutrinos $\nu_{Li}$, transform non-trivially, i.e., 
\begin{equation}
e_{Li} \to (T_l)_{ij}\,e_{Lj}\,, \qquad 
\nu_{Li} \to (S_\nu)_{ij}\,\nu_{Lj}\,.
\label{eq:residual_trans}
\end{equation}
Here, $T_l$ and $S_\nu$ are $3 \times 3$ unitary matrices that represent the underlying symmetry. As a result of this, the neutrino and charged lepton matrices, namely $M_\nu$ and $M_l$, satisfy\footnote{The mass matrices are defined in the basis, $(M_l)_{ij} \overline{e}_{Li} e_{Rj}$ and $(M_\nu^*)_{ij} \overline{\nu}_{Li} \nu_{Rj}$ for the Dirac neutrinos, and $\frac{1}{2}(M_\nu)_{ij} \overline{\nu}^T_{Li} C \nu_{Lj}$ for the Majorana neutrinos.}
\begin{equation}
S_\nu^{T} M_\nu M_\nu^\dagger S^*_\nu = M_\nu M_\nu^\dagger, \qquad 
T_\ell^{\dagger} M_\ell M_\ell^{\dagger} T_\ell = M_\ell M_\ell^{\dagger}.
\label{eq:residual}
\end{equation}
Additionally, $M_\nu$ satisfies a more restrictive criterion,
\begin{equation}
S_\nu^{T} M_\nu S_\nu = M_\nu\,,
\label{eq:residual_maj}
\end{equation}
If neutrinos are Majorana and possess lepton-number-violating mass terms.

It is a well-known result \cite{Lam:2007qc,Lam:2008rs,Lam:2008sh,Lam:2012ga,Lam:2011ag} that the absolute values of all or some of the elements of $U_{\rm PMNS}$ can be systematically inferred from Eq. (\ref{eq:residual}). The fact that $S_\nu$ $(T_l)$ commutes with $M_\nu M_\nu^\dagger$ $(M_l M_l^\dagger)$, the unitary matrix $U_\nu$ ($U_l$) that diagonalizes $M_\nu M_\nu^\dagger$ $(M_l M_l^\dagger)$ is related to $V_\nu$ $(V_l)$ which diagonalizes the symmetry generator $S_\nu$ $(T_l)$. One can then construct $U_{\rm PMNS} = U_l^\dagger U_\nu$. Following this strategy, several systematic and comprehensive investigations have been carried out, assuming that the residual symmetry groups $G_\nu$ and $G_l$ generated by $S_\nu$ and $T_l$, respectively, are subgroups of some bigger unifying group, $G_f$ which is the symmetry of leptonic interactions in the $SU(3)_C \times SU(2)_L \times U(1)_Y$ phase of the theory. The candidate $G_f$ as the DSG of $SU(3)$ \cite{Grimus:2010ak,Grimus:2011fk,Grimus:2013apa} or the DSG of $U(3)$ which is not the DSG of $SU(3)$ \cite{Ludl:2010bj,Joshipura:2014pqa,Joshipura:2015zla} has been extensively investigated for their predictions for entire or partial $U_{\rm PMNS}$. The aforementioned residual symmetries can also arise in modular invariant theories  when the value of the modulus or moduli is restricted to one of the self-dual or fixed points \cite{Novichkov:2018yse,Novichkov:2021evw,Novichkov:2022xrp,Ishiguro:2022pde,Hoshiya:2022qvr,Novichkov:2022fcg,Kashav:2024lkr}.

It turns out that determining the entire $U_{\rm PMNS}$ within the $3\sigma$ range of its observed values using the aforementioned residual-symmetry-based approach is not possible using reasonably small $G_f$. The main difficulty arises due to the considerably precise measurement of $|U_{e3}|$ in the last decade. The groups that contain $G_l$ and $G_\nu$ suitable to reproduce viable $|U_{e3}|$ along with the other elements were found to be of order $> 600$ \cite{Holthausen:2012wt}. While taking $G_f$ to be a large group is not technically disfavoured, it is generally regarded as a less well-motivated choice, given the landscapes of residual symmetries it allows. Selecting any particular residual symmetry then becomes essentially a parametric assumption rather than a consequence of the framework.

A less ambitious but more realistic approach is to predict only one of the three columns of $U_{\rm PMNS}$ group-theoretically, which is the main focus of the present investigation. In this case, one assumes $G_\nu = Z_2$ and $G_l = Z_m$ with $m\geq3$ for the residual symmetries \cite{Joshipura:2016quv}. The corresponding generators satisfy
\begin{equation}
S_\nu^{\,2}=\mathbb{I}, \qquad T_l^{\,m}=\mathbb{I}.
\end{equation}
Since the eigenvalues of $S_\nu$ can be $\pm 1$, it has at least one distinct eigenvalue, and the corresponding eigenvector is uniquely determined \footnote{Note that we do not consider a trivial case corresponding to $S_\nu = -\mathbb{I}$.}. We identify such an eigenvector of $S_\nu$ as a column $c_\nu$. Similarly, choosing $T_l$ such that all its eigenvalues are distinct, the unitary matrix $V_l$ can be determined through $V_l^{\dagger} T_l V_l = d_l$ with $d_l$ diagonal and real. Subsequently, 
\begin{equation}
c_0 = V_l^{\dagger} c_\nu\,,
\end{equation}
represent a column fixed entirely and uniquely by the residual symmetries. The absolute values of the elements of $c_0$ can be considered as predictions for the absolute values of a column in $U_{\rm PMNS}$ up to freedom of permutation of the elements of $|c_0|$. The latter arises from the fact that the residual symmetries do not predict the masses of the charged leptons.

\section{Additional Plots}\label{app:plots}
This appendix presents supplementary results for the fixed-column predictions $C2[1]$, $C2[2]$, $C1[6](C1'[6])$ and $C1[7](C1'[7])$, which are not shown in the main text. Although these columns are disfavoured by the latest global fit in view of their predictions for $\theta_{12}$, we proceed to analyse the associated $\sin^2\theta_{12}$--$\delta_{CP}$ correlations. The corresponding figures display the allowed regions in the $\sin^2\theta_{23}$--$\delta_{\rm CP}$ plane obtained from DUNE and Hyper-K, both individually and in combination, with and without the inclusion of the JUNO prior on $\sin^2\theta_{12}$. $C1[6](C1'[6])$ and $C1[7](C1'[7])$ exhibit particularly strong constraints, with large portions of the parameter space excluded, thereby illustrating the enhanced discriminatory power of precision measurements when combined across experiments.
%%%%%%%%%%%%%%%%%%%%%
\begin{figure}[!h]
    \centering
\includegraphics[width=0.49\linewidth]{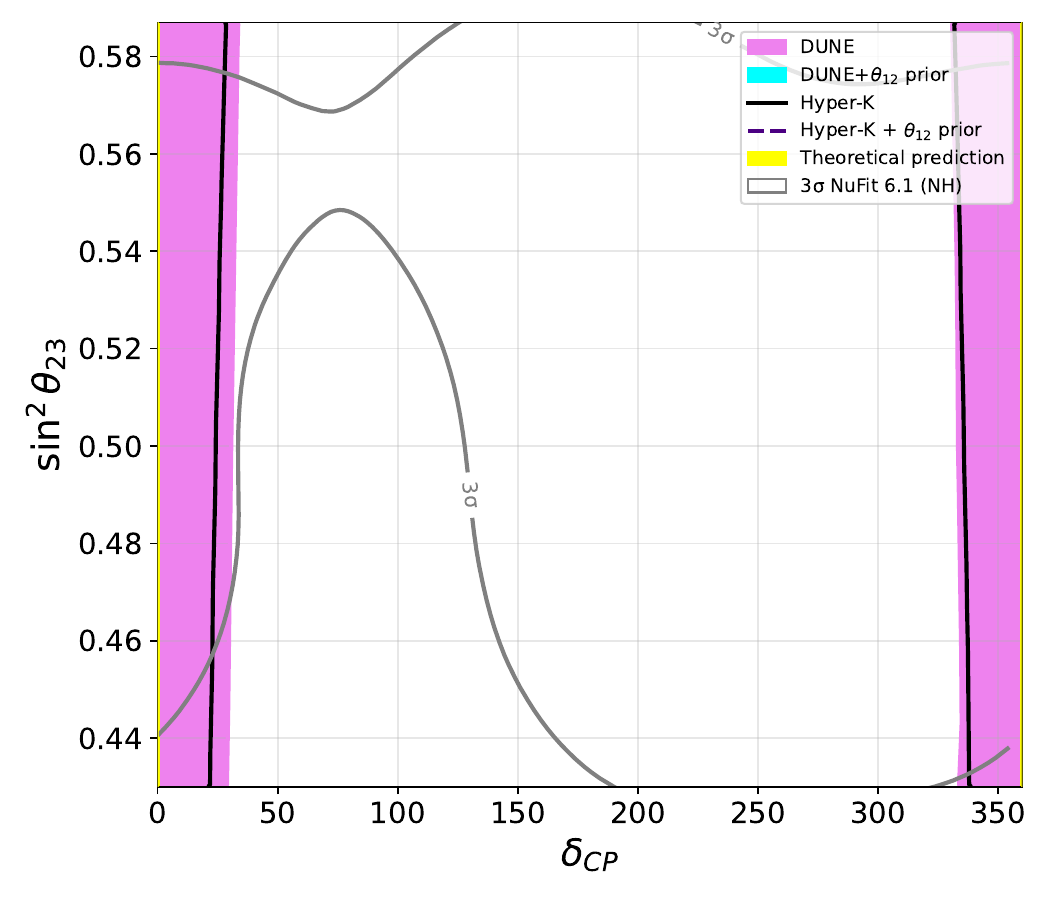}
\includegraphics[width=0.49\linewidth]{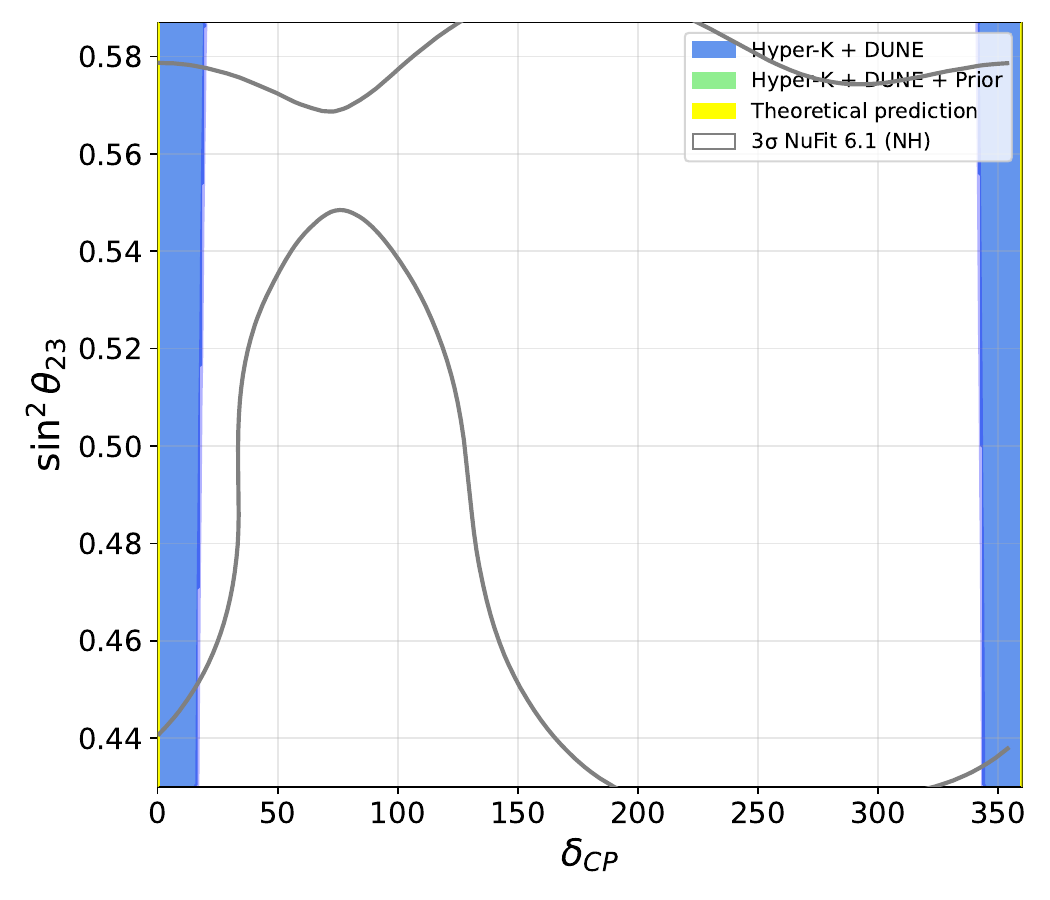}
\includegraphics[width=0.49\linewidth]{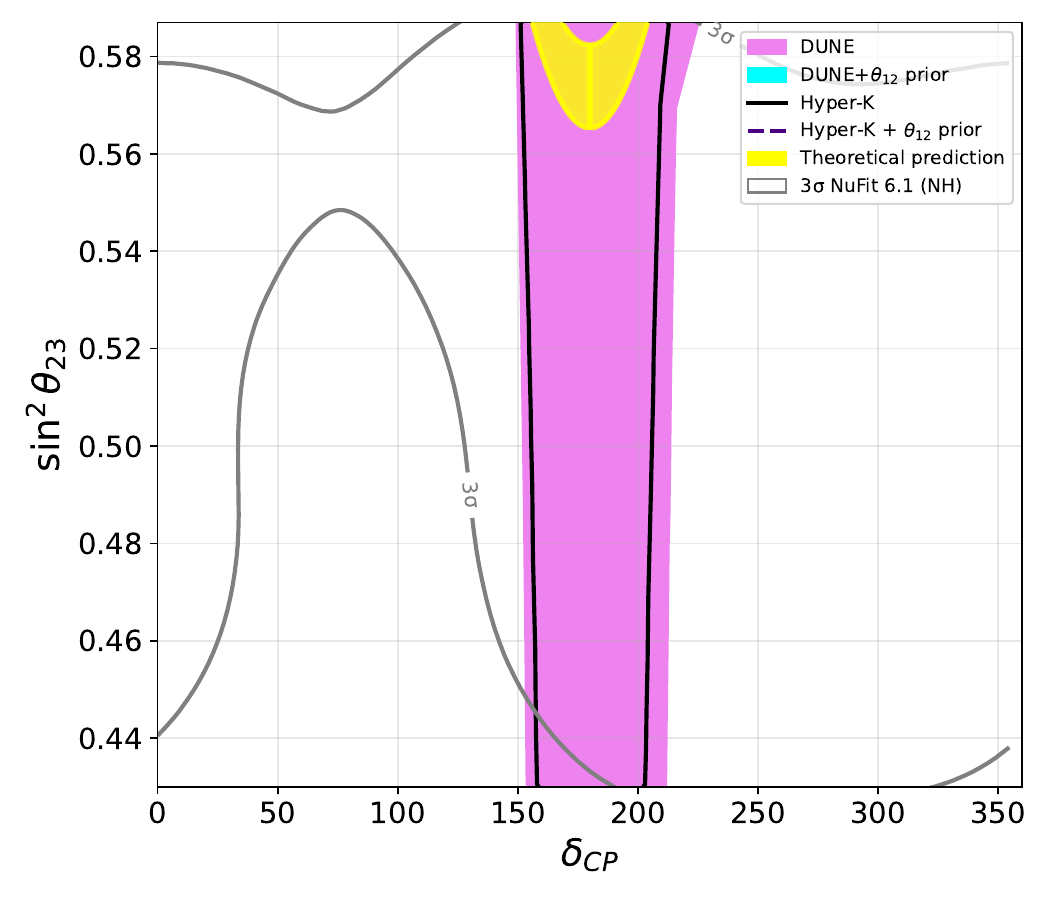}
\includegraphics[width=0.49\linewidth]{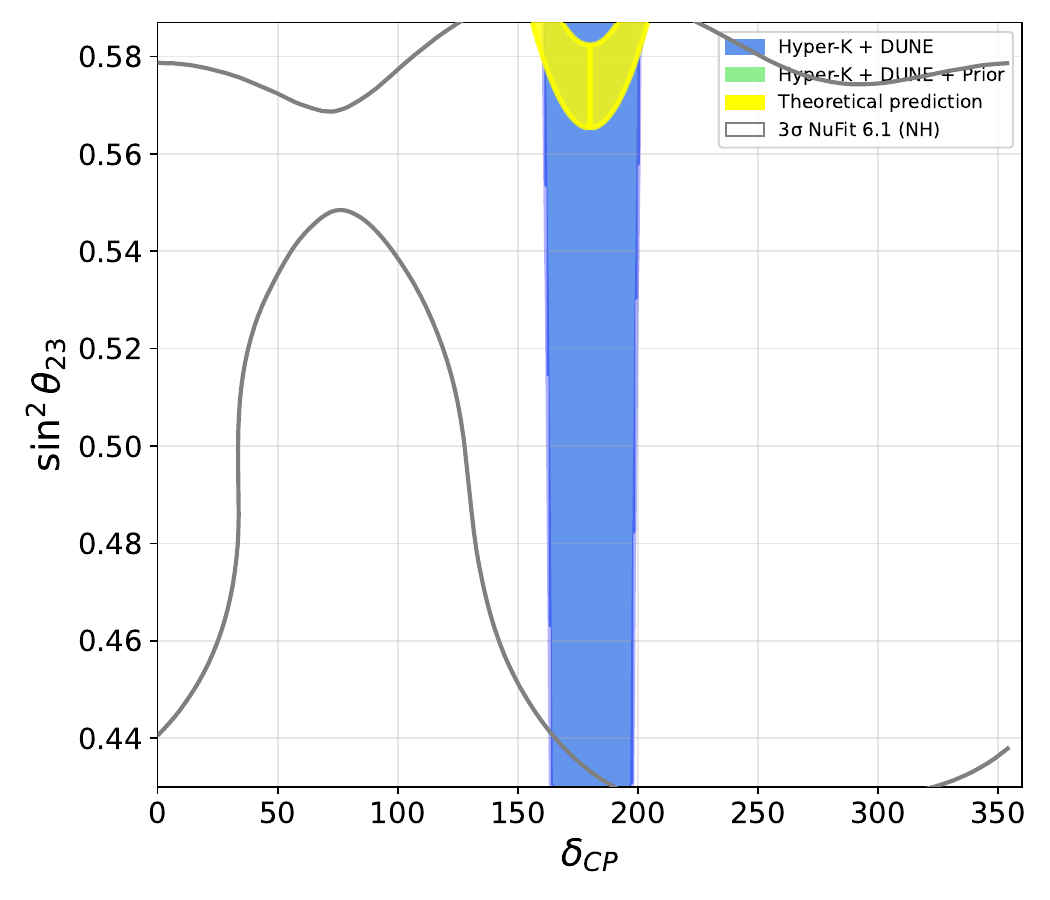}
\caption{Allowed regions in the $\sin^2\theta_{23}$--$\delta_{\rm CP}$ plane for the fixed column predictions $C1[6]$ (upper) and $C1'[6]$ (lower) from Table~\ref{tab:1}. The same experimental sensitivities and theoretical correlations as in Fig.~\ref{fig1} are used.}
\label{fig8}
\end{figure}
\begin{figure}[!h]
    \centering
\includegraphics[width=0.49\linewidth]{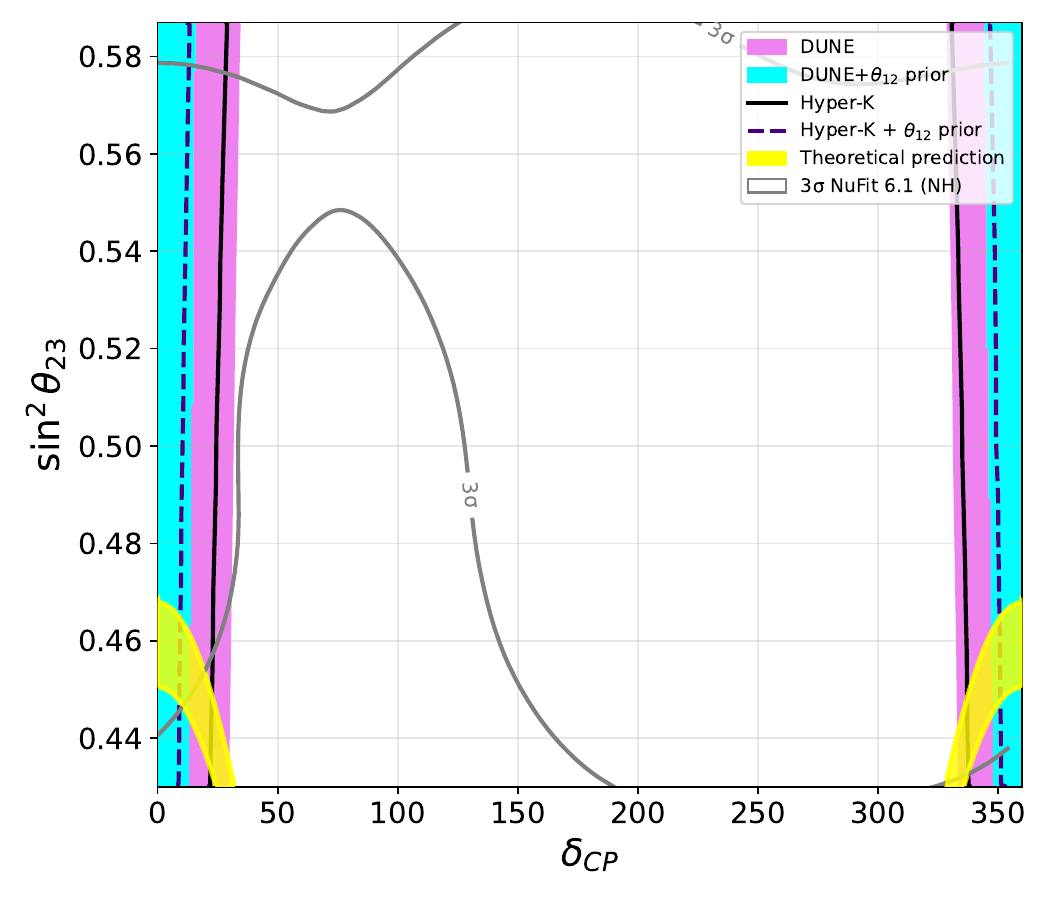}
\includegraphics[width=0.49\linewidth]{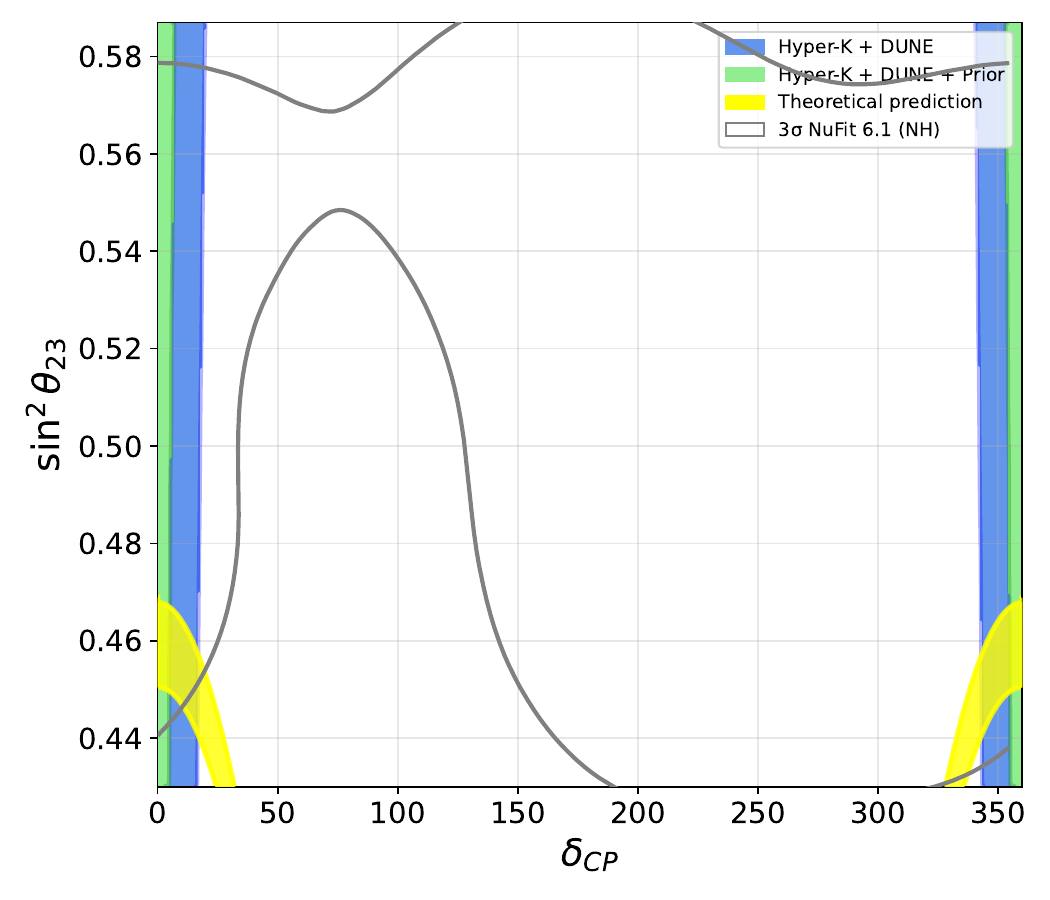}
\includegraphics[width=0.49\linewidth]{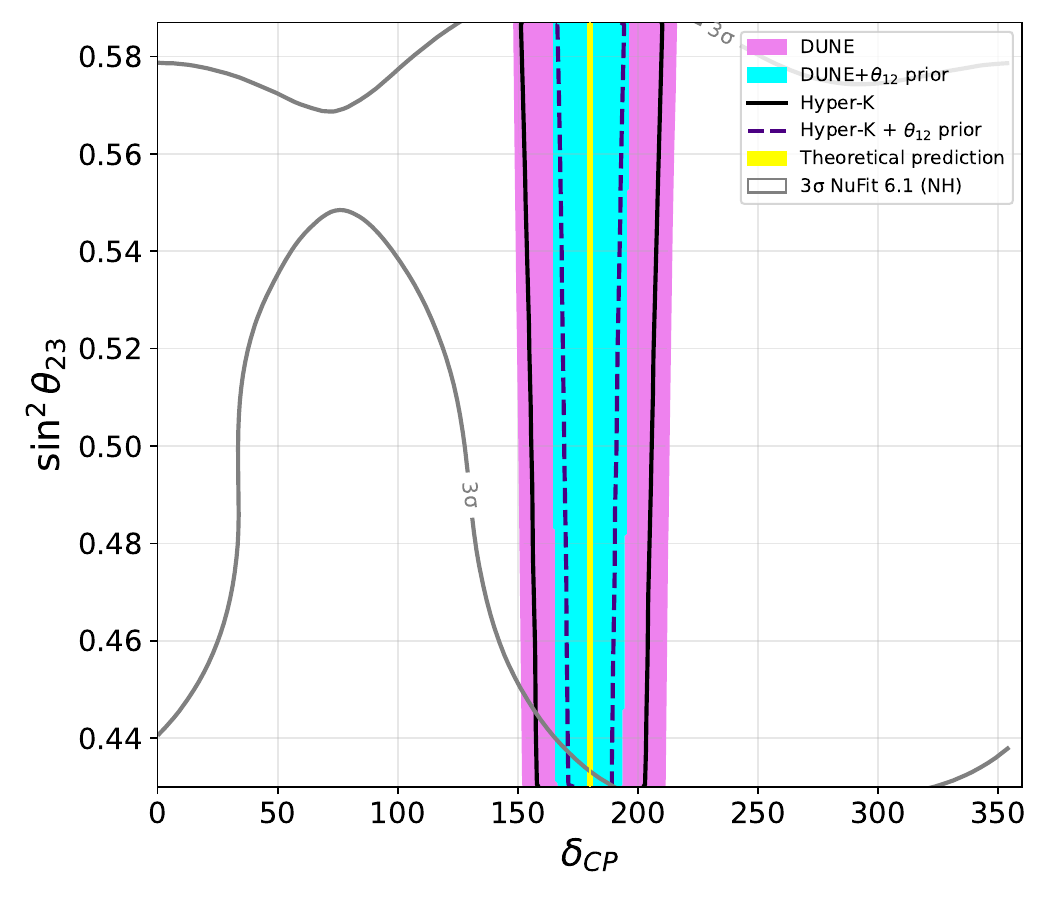}
\includegraphics[width=0.49\linewidth]{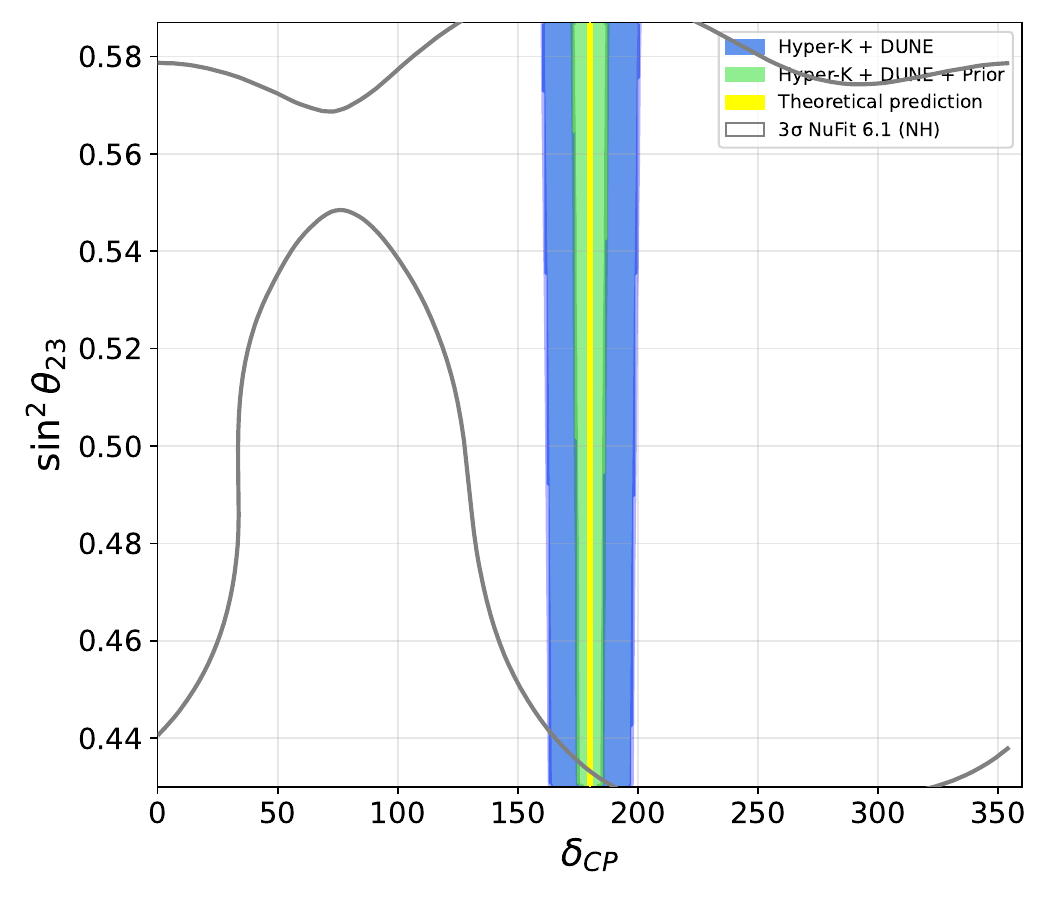}
\caption{Allowed regions in the $\sin^2\theta_{23}$--$\delta_{\rm CP}$ plane for the fixed column prediction $C1[7]$ (upper) and $C1'[7]$ (lower)  from Table~\ref{tab:1}, shown using the same experimental and theoretical conventions as in Fig.~\ref{fig1}.}
\label{fig9}
\end{figure}
%%%%%%%%%%%%%%%%%%%%

\begin{figure}[!htbp]
\centering
\includegraphics[width=0.49\linewidth]{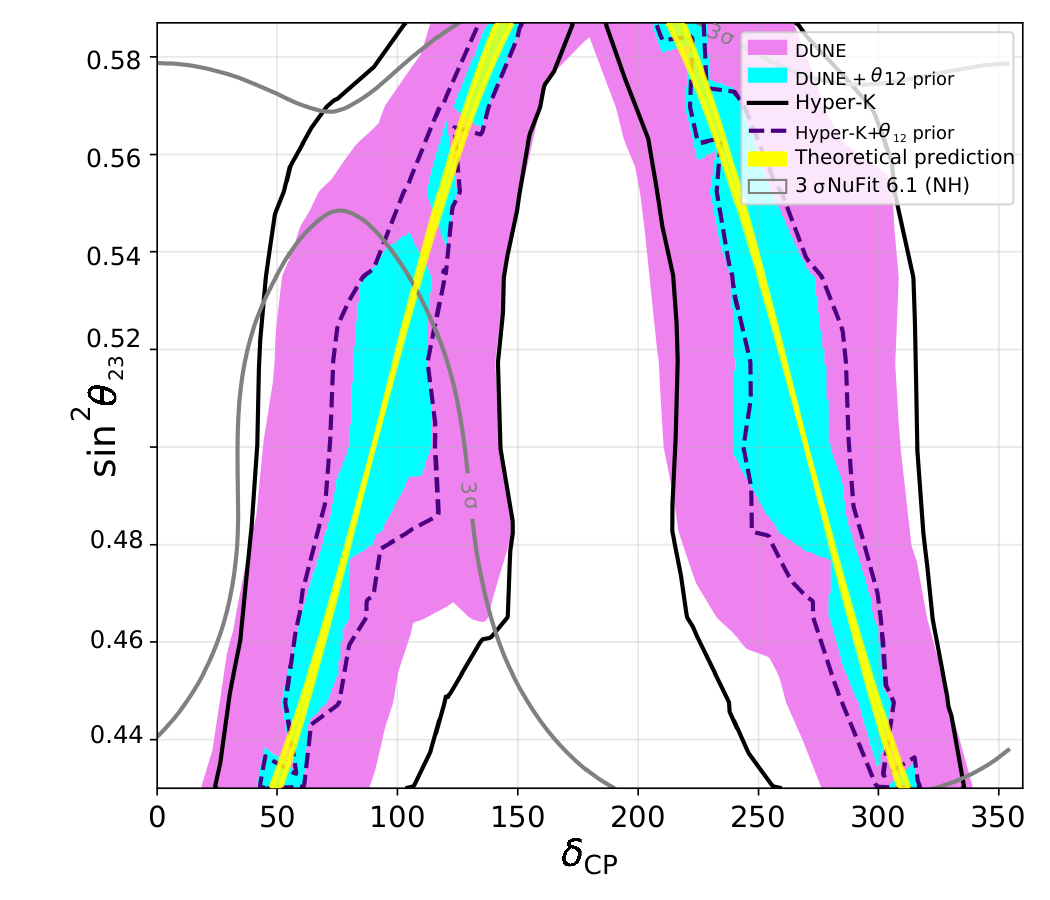}  
\includegraphics[width=0.49\linewidth]{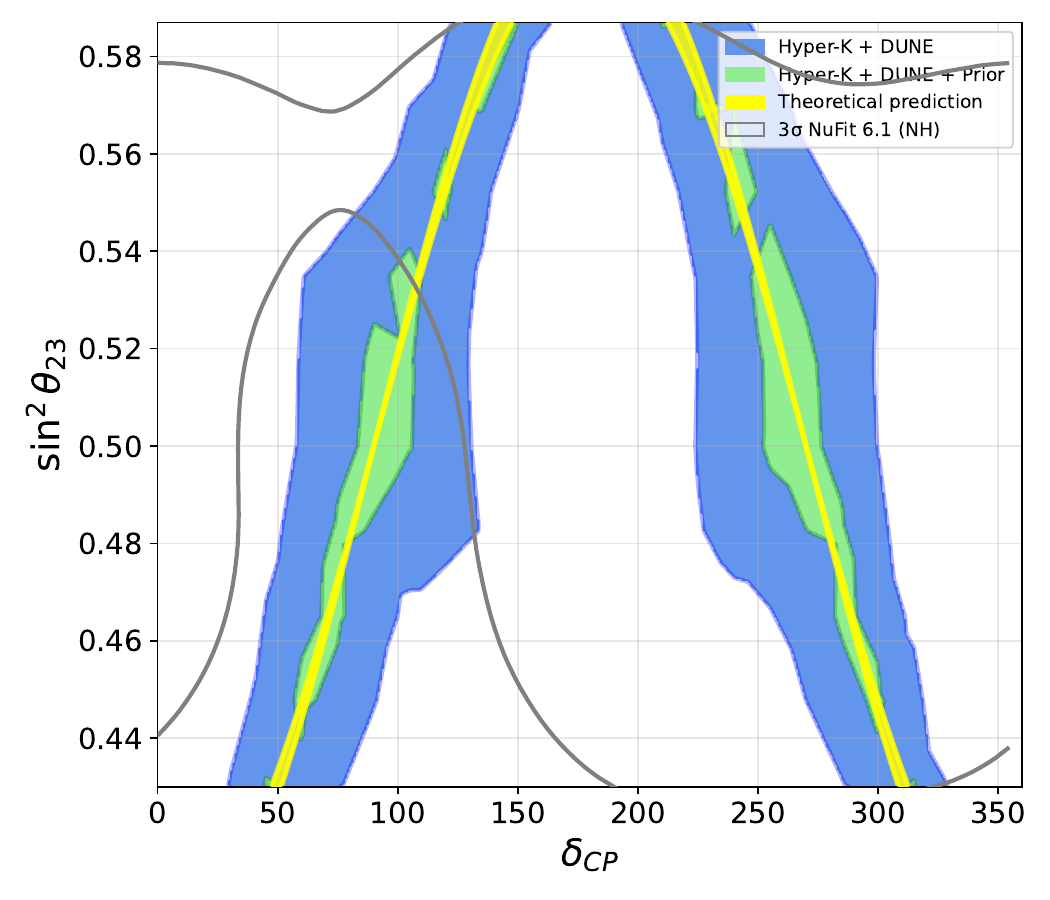}
\includegraphics[width=0.49\linewidth]{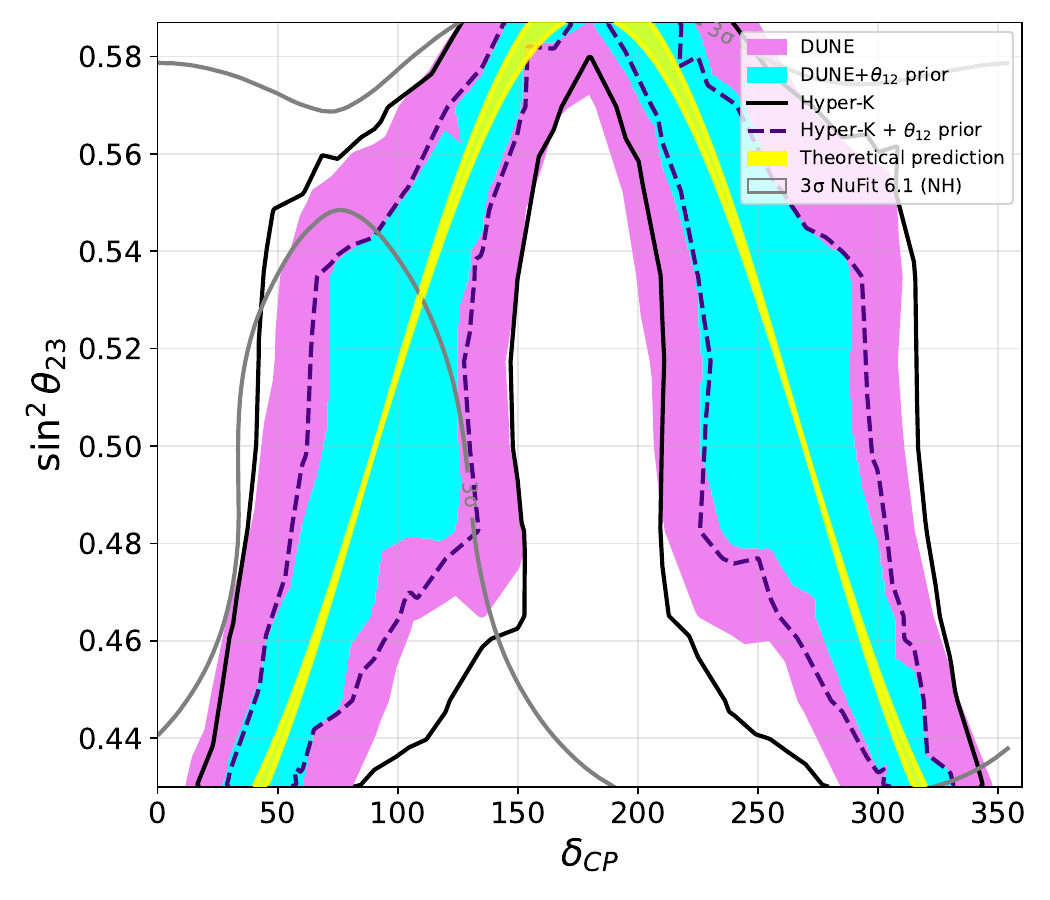}
\includegraphics[width=0.49\linewidth]{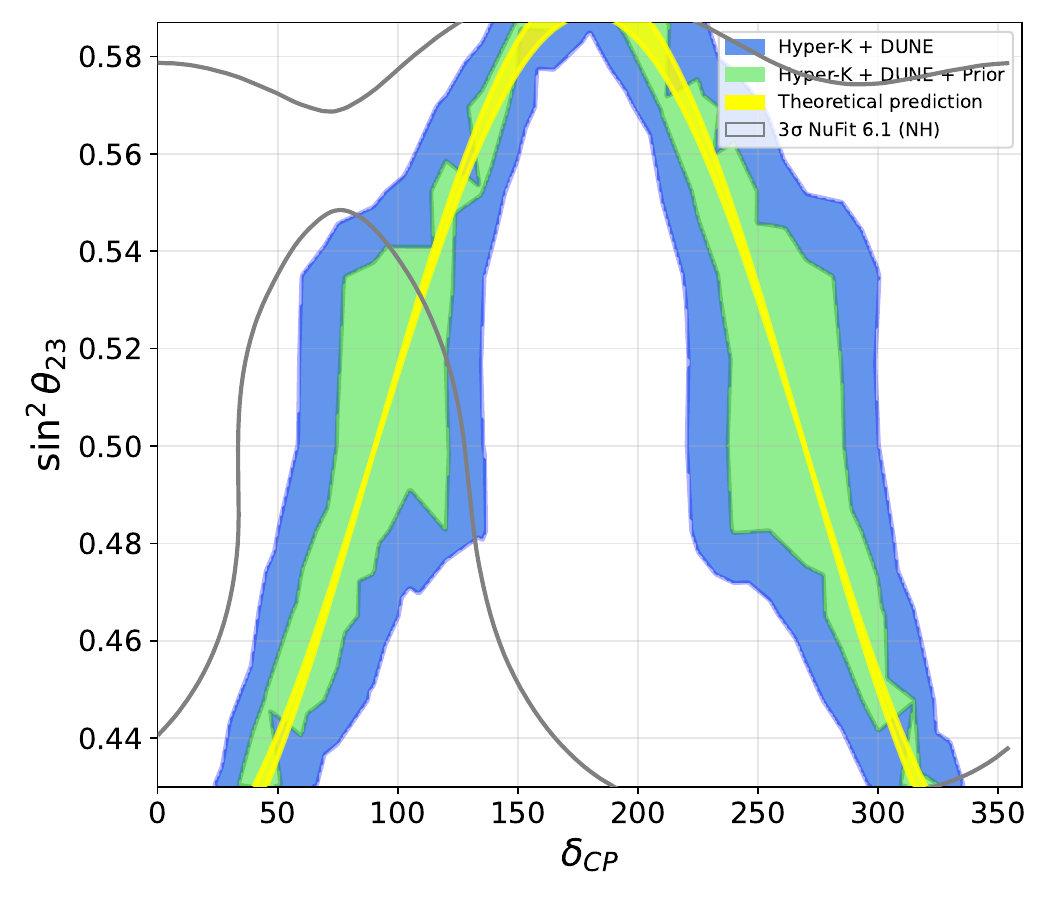}
\caption{Allowed regions in the $\sin^2\theta_{23}$--$\delta_{\rm CP}$ plane for the fixed column prediction $C2[1]$ (upper) and  $C2[2]$ (lower) from Table~\ref{tab:1}, shown using the same experimental and theoretical conventions as in Fig.~\ref{fig1}.
}
\label{fig10}
\end{figure}

For the $C2[1]$ and $C2[2]$ cases, inclusion of the $\sin^2\theta_{12}$ prior leads to a noticeable upward shift in the corresponding $\chi^2_{\min}$. While $\chi^2_{\min}$ remains below the $3\sigma$ threshold of $11.83$, the enlarged minimum values cause a substantial contraction of the cyan and green allowed regions, clearly reflecting the impact of the prior. An analogous trend is observed in the results displayed in Fig.~\ref{fig10}.
%%%%%%%%%%%%%%%%%%%%%%%%%%%%
%%%%%%%%%%%%%%%%%%%%%%%%%%

\bibliographystyle{JHEP}
%\bibstyle{apsrev}
\bibliography{newref1}

\end{document}